\documentclass[conference]{svproc}  
\usepackage{times}

\usepackage{graphics} 
\usepackage{epsfig} 
\usepackage{mathptmx} 
\usepackage{times} 
\usepackage{amsmath} 
\usepackage{amssymb}  
\usepackage{bm}  
\usepackage{newtxtext, newtxmath}
\usepackage{color}
\usepackage{cite}
\usepackage[font=footnotesize]{caption}
\usepackage{subcaption}
\usepackage{mathtools}
\usepackage{lscape}
\usepackage[graphicx]{realboxes}
\newenvironment{changemargin}[2]{%
\begin{list}{}{%
\setlength{\topsep}{0pt}%
\setlength{\leftmargin}{#1}%
\setlength{\rightmargin}{#2}%
\setlength{\listparindent}{\parindent}%
\setlength{\itemindent}{\parindent}%
\setlength{\parsep}{\parskip}%
}%
\item[]}{\end{list}}

\usepackage{algpseudocode}

\algnewcommand{\IfThenElse}[3]{
  \State \algorithmicif\ #1\ \algorithmicthen\ #2\ \algorithmicelse\ #3}

\newtheorem{assumption}{Assumption}

\usepackage{algorithm}
\usepackage{algpseudocode}

\algnewcommand{\IfThen}[2]{
  \State \algorithmicif\ #1\ \algorithmicthen\ #2}
  
\makeatletter
\let\OldStatex\Statex
\renewcommand{\Statex}[1][3]{%
  \setlength\@tempdima{\algorithmicindent}%
  \OldStatex\hskip\dimexpr#1\@tempdima\relax}
\makeatother

\newcommand{\reals}{\mathbb{R}}

\newcommand{\pstate}{p}
\newcommand{\tstate}{s}
\newcommand{\rstate}{r}
\newcommand{\rt}{\tilde{\rstate}}
\newcommand{\estate}{e}
\newcommand{\astate}{\eta}
\newcommand{\rtrans}{\phi}
\newcommand{\dyn}{f}
\newcommand{\ctrl}{u}

\newcommand{\pdyn}{\dyn^\pstate}
\newcommand{\tdyn}{\dyn^\tstate}
\newcommand{\rdyn}{\dyn^\rstate}

\newcommand{\pctrl}{\ctrl^\pstate}
\newcommand{\tctrl}{\ctrl^\tstate}

\newcommand{\pcset}{\mathcal U^\pstate}
\newcommand{\tcset}{\mathcal U^\tstate}

\newcommand{\eye}{\mathbf I}
\newcommand{\zero}{\mathbf 0}

\newcommand{\valfunc}{V} 
\newcommand{\teb}{\mathcal B}
\newcommand{\vf}{V} 
\newcommand{\vft}{\tilde{V}}
\newcommand{\elpsf}{E}
\newcommand{\elps}{\mathcal\elpsf} 

\newcommand{\lag}{L}
\newcommand{\laglya}{\lag^\text{Lya}}
\newcommand{\lagpc}{\lag^p} 
\newcommand{\lagtc}{\lag^s}

\newcommand{\lagelps}{\lag^\elps}

\newcommand{\ptmat}{Q} 

\newcommand{\issos}{\text{ is SOS}}
\newcommand{\setfunc}{g}
\newcommand{\fbctrl}{K}

\renewcommand{\baselinestretch}{0.9}

\newcommand{\revision}[1]{{\color{black}{#1}}}

\title{Robust Tracking with Model Mismatch for Fast and Safe Planning: an SOS Optimization Approach\vspace{-.5em}
}

\author{Sumeet Singh\inst{1} \and Mo Chen\inst{1}\and Sylvia L. Herbert\inst{2}\and Claire J. Tomlin\inst{2}\and Marco Pavone\inst{1}\thanks{Singh, Chen, and Pavone were supported by NASA under the Space Technology
Research Grants Program, Grant NNX12AQ43G, and by the King Abdulaziz City for Science and Technology (KACST). Herbert and Tomlin were supported by SRC under the CONIX Center and by ONR under the BRC program in Multibody Systems.
}
}

\institute{Dept. of Aeronautics and Astronautics, Stanford University\\ \texttt{\{ssingh19, mochen72, pavone\}@stanford.edu} 
\and
Dept. of Electrical Engineering and Computer Science, University of California, Berkeley\\
\texttt{\{sylvia.herbert, tomlin\}@berkeley.edu}\vspace{-2em}}

\begin{document}

\maketitle
\thispagestyle{empty}
\pagestyle{empty}

\begin{abstract}
In the pursuit of real-time motion planning, a commonly adopted practice is to compute trajectories by
running a planning algorithm on a simplified, low-dimensional dynamical model, and then employ a feedback tracking controller that tracks such a trajectory by accounting for the full, high-dimensional  system dynamics. While this strategy of \emph{planning with model mismatch} generally yields fast computation times, there are no guarantees of dynamic feasibility, which hampers application to safety-critical systems. Building upon recent work that addressed this problem through the lens of Hamilton-Jacobi (HJ) reachability, we devise an algorithmic framework whereby one computes, offline, for a pair of ``planner" (i.e., low-dimensional) and ``tracking" (i.e., high-dimensional) models, a feedback tracking controller and associated   tracking bound. This bound is then used as a safety margin when generating motion plans via the low-dimensional model. Specifically, we harness the computational tool of sum-of-squares (SOS) programming to design a  bilinear optimization algorithm for the computation of the feedback tracking controller and associated tracking bound. The algorithm is demonstrated via numerical experiments, with an emphasis on investigating  the trade-off between the increased computational scalability afforded by SOS and its intrinsic conservativeness. Collectively, our results enable scaling the appealing strategy of  planning with model mismatch to systems that are beyond the reach of HJ analysis, while maintaining safety guarantees. \vspace{-1em}
\end{abstract}

\section{Introduction} 
\begin{figure}[t]
	\centering
	\includegraphics[width=0.27\textwidth]{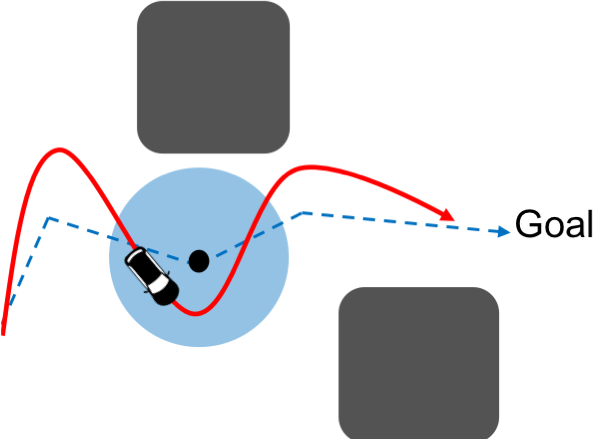} \qquad \qquad
  \includegraphics[width=0.48\textwidth]{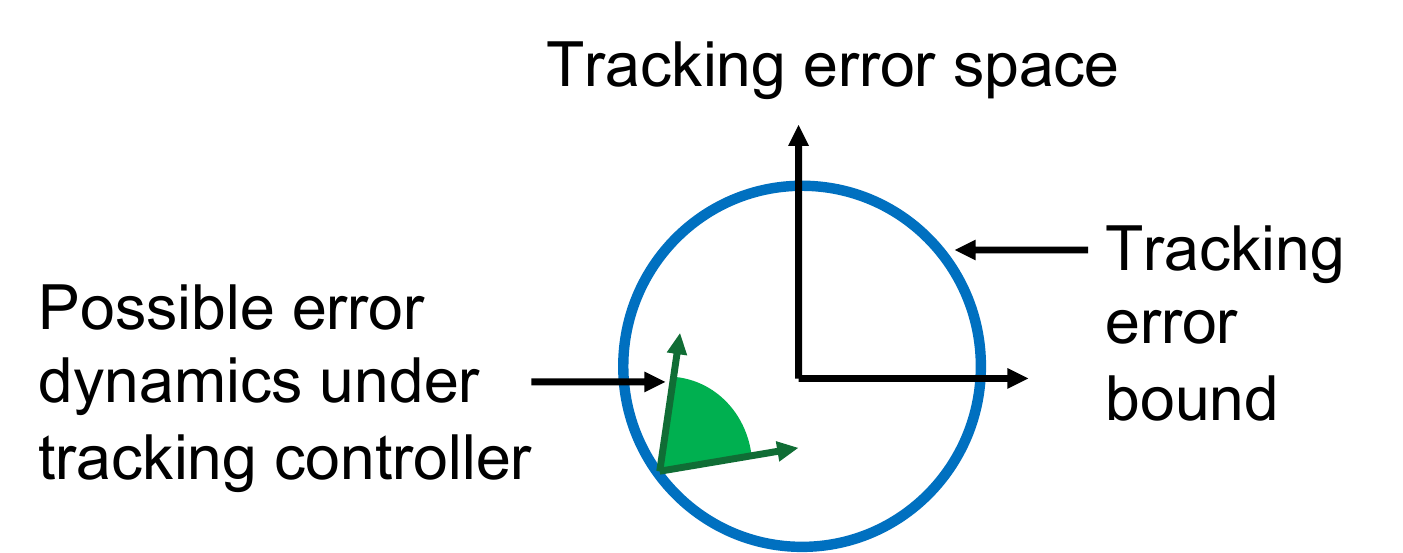}
	\caption{{\em Left: Planning with model mismatch.} The tracking system (car, red trajectory) follows a motion plan computed via a low-fidelity dynamical model (blue trajectory). 
Tracking with model mismatch while maintaining safety requires keeping the maximum tracking error bound (TEB) below a certifiable value.
{\em Right: TEB certification.} The TEB is characterized by the property that on its boundary, all possible error dynamics resulting from the nominal trajectory point inwards under some tracking controller, so that the TEB is never exceeded.	\vspace{-2.1em}}
	\label{fig:intropic}
\end{figure}

\vspace{-.7em}
As robotic systems become more pervasive, real-time motion planning becomes increasingly important. In general, motion planning algorithms must find a trade-off among three key challenges: (1) achieving fast computational speed to enable online re-planning, (2) accounting for complex system dynamics, and (3) ensuring formal safety and robustness guarantees.  In particular,  to enable high-frequency re-planning, a commonly adopted practice is to compute a trajectory by
running a planning algorithm on a simplified, low-dimensional dynamical model, oftentimes just a single-integrator model. A feedback tracking controller that accounts for the full, high-dimensional dynamics of the system is then used to track such a trajectory (we refer to this approach as {\em planning with model mismatch}, see Figure \ref{fig:intropic}). 
This approach usually guarantees fast planning speeds, but guaranteeing safety becomes difficult, due to the mismatch between the model used for planning and the actual dynamics of the robotic system.
This approximation can introduce tracking error between the planned path and the actual trajectory of the system, potentially resulting in safety violations. 
Conversely, one could compute a collision-free motion plan by directly  accounting for the full set of differential constraints of a robotic system (e.g., by using a general-purpose kinodynamic motion planning algorithm \cite{LaValle2011}). However, despite significant recent progress \cite{LaValleKuffner2001,KaramanFrazzoli2011,JansonSchmerlingEtAl2015,LiLittlefieldEtAl2016}, kinodynamic motion planning is still computationally intensive \cite{LaValle2011}. For robots characterized by ``slow" or differentially flat dynamics, gradient-based trajectory optimization techniques (e.g.,\cite{RatliffZuckerEtAl2009,SchulmanHoEtAl2013}) may provide quick convergence to  a {feasible} solution, but extending such methods to systems with complex dynamics is challenging.

In general, providing formal guarantees on safety and robustness typically requires reasoning on the full dynamical model of a robotic system. One of the most powerful tools in this regard is the concept of  ``funnels'' \cite{BurridgeRizziEtAl1999}. The key idea is to ensure that the trajectory of a system remains within precomputed funnels \cite{TedrakeManchesterEtAl2010, MajumdarTedrake2013, MajumdarAhmadiEtAl2013}; this is often achieved via Lyapunov analysis techniques. However, while able to provide formal guarantees around nominal trajectories, these methods are less suitable for \textit{a priori} unknown environments in which safe trajectories must be found online. The work in \cite{MajumdarTedrake2017} tackles this challenge by considering a library of funnels, from which maneuvers are composed in sequence to form guaranteed safe trajectories. To obtain trajectory-independent tracking error bounds, contraction theory and nonlinear optimization techniques can be used to construct tubes around {\em dynamically-feasible} trajectories obtained from any planning algorithm \cite{SinghMajumdarEtAl2017}. However, as mentioned above, rapidly computing {dynamically-feasible} nominal trajectories poses a significant computational challenge.

To tackle real-time motion planning problems, recent works such as \cite{ChenBansalEtAl2018, HerbertChenEtAl2017, KousikVaskovEtAl2017} combine low- and high-fidelity models to strike a balance among the three aforementioned challenges of computational speed, model accuracy, and robustness.
In \cite{KousikVaskovEtAl2017}, a forward reachable set for the high-fidelity model is computed offline and then used to prune trajectories generated online using the low-fidelity model. The approach   relies on an {\em assumed} model mismatch bound, expressed as the maximum distance between the low- and high-fidelity models under a certain metric.
FaSTrack \cite{ChenBansalEtAl2018, HerbertChenEtAl2017} considers a slightly different definition of model mismatch, and casts a motion planning problem as a differential game wherein a low-fidelity ``planning" system is being pursued (i.e., tracked) by a  high-fidelity ``tracking" system. Hamilton-Jacobi (HJ) reachability analysis is used to precompute a worst-case tracking error bound (TEB) as well as the optimal feedback tracking controller. 
Online, FaSTrack plans in real time using the low-fidelity planning system, while simultaneously tracking the plan and remaining within {\em guaranteed} tracking error bounds -- despite the model mismatch. 
FaSTrack's ability to ensure real-time planning of dynamic systems with safety guarantees makes it a desirable framework, but it requires a significant precomputation step in computing the TEB and controller.
Executing this precomputation using HJ reachability is reliable and straightforward for nonlinear, highly-coupled dynamical systems with up to five states, or higher-dimensional systems of a specific form \cite{ChenHerbertEtAl2018}.
However, the computations required for HJ reachability analysis scale exponentially with the number of states, thus making FaSTrack difficult to apply to systems with a large number of states and/or highly-coupled dynamics.


Building upon FaSTrack, we present an approach to designing a feedback tracking controller along with guaranteed tracking error bounds via sum-of-squares (SOS) programming. SOS programs can be cast as semidefinite programs (SDPs), which may be solved using standard convex optimization toolboxes. SOS programming has been used extensively to provide convex relaxations to fundamentally non-convex or even NP-hard problems arising in robust control \cite{TedrakeManchesterEtAl2010, MajumdarTedrake2013, MajumdarAhmadiEtAl2013, MajumdarTedrake2017,PosaKoolenEtAl2017}.
In the context of this work, we leverage SOS to derive \emph{sufficient} conditions for bounding tracking control error. The price paid for such a convex relaxation is in the tightness of the error bound guarantee (which corresponds to the size/computational tractability of the SOS SDP), but critically not in the existence of the guarantee itself provided the SOS program is feasible.  SOS programming thus represents an attractive alternative to exact HJ reachability for those robotic systems where HJ is not directly applicable due computational intractability. In addition, this paper, as well as FaSTrack \cite{ChenBansalEtAl2018, HerbertChenEtAl2017}, has potential to complement works such as \cite{KousikVaskovEtAl2017} that assume a model-mismatch error, by providing the TEB as well as a feedback tracking controller to achieve such a TEB.


\emph{Statement of Contributions}: The contribution of this paper is threefold. First, we provide an SOS-based formulation for tracking a nominal motion plan computed on a simplified low-dimensional model.
Second, leveraging the SOS formulation, we devise a bilinear optimization algorithm to jointly compute, offline, a \textit{trajectory-independent} feedback tracking controller and the corresponding TEB. Finally, we demonstrate our approach on a 5D example for comparison with FaSTrack \cite{HerbertChenEtAl2017}, and on a 8D example which is beyond the reach of HJ analysis. Collectively, our three contributions enable scaling the appealing strategy of  planning with model mismatch to systems that are beyond the reach of HJ analysis, while maintaining safety guarantees.

\emph{Organization}: The remainder of the paper is organized as follows. In Section~\ref{sec:bg} we provide the problem statement and give necessary background on SOS programming. Section~\ref{sec:sos} formalizes the problem as a bilinear SOS program, while Section~\ref{sec:solve_sos} develops an algorithmic framework for its solution. In Section~\ref{sec:num} we present numerical experiments. Finally, in Section~\ref{sec:conclusion}, we draw our conclusions and suggest directions for future work.\vspace{-1em}

\section{Preliminaries}\label{sec:bg} \vspace{-1em}
In this section, we present the problem formulation, and provide the necessary background material related to sum-of-squares programming.\vspace{-1em}
\subsection{Problem Formulation}\label{subsec:form} 
Consider a relatively high-fidelity model of a robotic system, referred to as the  ``tracking model,'' and a lower-fidelity model, referred to as the ``planning model,'' described, respectively, by the ordinary differential equations (ODEs) \vspace{-1.5em}

\begin{align}
\dot\tstate = \tdyn(\tstate, \tctrl), \quad \tctrl \in \tcset, && \dot\pstate = \pdyn(\pstate, \pctrl), \quad \pctrl \in \pcset,
\label{tracker}
\end{align}

\noindent where $\tstate \in \reals^{n_s}$ is the tracking model's state (or ``tracking state''), and $\tctrl \in \reals^{m_s}$ is its control input (or ``tracking control''), constrained to lie within the compact set $\tcset$. Similarly, $\pstate \in \reals^{n_p}$ is the planning state and $\pctrl \in \reals^{m_p}$ is the planning control input, constrained to lie within the compact set $\pcset$.  The planning model can be thought of as a coarser or lower-fidelity model for which motion planning can be done effectively in real time. 
For both ODEs in equation \eqref{tracker}, we assume that the dynamics are Lipschitz continuous in the state for fixed controls, so that given any measurable control function, a unique trajectory exists for each system \cite{CoddingtonLevinson1955}. Furthermore, we assume that {\em the planning state $\pstate$ is a strict subset of the tracking state $\tstate$}, in particular $n_p<n_s$ (this assumption allows one to clearly relate the tracking and planning state, as discussed in Section \ref{subsec:Relsys}), and that the dynamics for the tracking model  are {\em control affine} (a rather mild condition that significantly simplifies the SOS formulation, as discussed in Section \ref{subsec:SOSconstraints}).

Crucially, trajectories generated by using the planning model may not be dynamically feasible with respect to the tracking model. Thus, a tracking controller that accounts for the high-fidelity dynamics of the robotic system, as represented by the tracking model, may not be able to track these trajectories exactly and some (possibly large) tracking error may be incurred.
Figure \ref{fig:intropic} illustrates this behavior, for the case of a 4D car model used to track  a 2D single integrator model.
Specifically, a planning algorithm is used to ``quickly'' find a nominal motion plan for the planning model, depicted by the dashed blue line. 
The robot, represented more accurately by the tracking model, strives to track the nominal motion plan as closely as possible, giving rise to the red trajectory. 
The presence of a minimum turning radius constraint in the tracking model prevents the robot from  tracking the sharp corners exactly, thereby yielding a tracking error.

Such a decoupling approach is quite common in the field of motion planning
\cite{LaValle2011}, yet very few results exist on how to account for the
resulting tracking error. In particular, we advocate that the computation of the nominal motion plan via the planning model should account for the tracking error in order to ensure safe execution (i.e., tracking) by the real system, described by the tracking model. Accordingly, the objective of this paper is to devise an efficient and scalable algorithm to compute (1) a \textit{trajectory-independent} TEB (represented by the light blue circle in Fig.~\ref{fig:intropic}, left), which the planner can use as a ``safety buffer" to ensure that the actual trajectory is collision free, and (2) a feedback tracking controller to track such a nominal motion plan while respecting the TEB. Intuitively, the tracking controller should have the property that on the boundary of the TEB, the error dynamics are driven ``inwards,'' \textit{regardless of the nominal motion plan} (depicted in Fig.~\ref{fig:intropic}, right). \vspace{-1em}

\subsection{Sum-of-Squares (SOS) Programming Background}
Our approach is rooted in SOS programming, for which we provide a brief review here (for a more detailed review of SOS programming and its applications, please refer to \cite{Parrilo2000,AhmadiMajumdar2016,MajumdarTedrake2017}). We begin by discussing semidefinite programs (SDPs), a class of convex optimization problems formulated over the space of symmetric positive semidefinite (psd) matrices. Denote the set of $n\times n$ symmetric matrices as $S^n$. A symmetric matrix $X \in S^n$ is positive semidefinite (psd) if $x^\top X x \geq 0$ for all $x\neq 0$, and is denoted as $X \succeq 0$. An SDP in standard form is written as:

\begin{equation}\label{eq:sosopt-general}
\begin{aligned}
& \underset{X\in S^n}{\text{minimize}}
& & \text{Tr} (C^\top X) \\
& \text{subject to}
& & A_i^\top X = b_i, \quad i = 1,2,...,m,\\
& & & X \succeq 0
\end{aligned}
\end{equation}

\noindent where $C$ and $\{A_i\}_i$ are elements of $S^n$, and ``Tr'' denotes the trace operator. SOS programs provide a means of certifying nonnegativity of polynomials either globally or over basic semialgebraic sets. A basic semialgebraic set $\mathcal{S}$ is a subset of the Euclidean space characterized by a finite set of polynomial inequalities and equalities, that is, \vspace{-.5em}
\begin{equation}
\mathcal{S} := \{ x \in \reals^n : \phi_i(x) \leq 0,\,  \psi_j(x) = 0 \},
\label{sa_set} \vspace{-.5em}
\end{equation}
where $\{\phi_i\}, \{\psi_j\}$ are multivariate polynomials in $x$. The simplest task within this class of problems is verifying the nonnegativity of a given polynomial $p(x)$ over $\mathcal{S}$, which is already NP-hard \cite{Parrilo2000}. SOS programming provides a convex relaxation approach to accomplish this task. Specifically, a polynomial $p$ is SOS if it can be written in the form $\sum_k z_k^2$ for some other polynomials $z_k(x)$. While such a decomposition is not necessary, it is sufficient to guarantee (global) nonnegativity of $p(x)$. If one can find a set of SOS polynomials $L_i(x)$ and ordinary polynomials $q_j(x)$ such that \vspace{-.9em}
\begin{equation}
p + \sum_i L_i \phi_i  + \sum_j q_j\psi_j \issos,
\label{sos_cert}\vspace{-.9em}
\end{equation}
then one obtains a {\em certificate} of nonnegativity of $p(x)$ over $\mathcal{S}$. Indeed, in the above equation, when $\phi_i(x) \leq 0$ and $\psi_j(x) = 0$, one has that $p(x) \geq -\sum_i L_i(x) \phi_i(x) \geq 0$, as required. Such a certificate is the extension of the generalized S-procedure~\cite{IwasakiHara2005} to the setting of real-valued polynomials~\cite{Parrilo2000}, and additionally constitutes a necessary condition for a subclass of semialgebraic sets~\cite{Putinar1993}. 

The computational advantage of SOS programming stems from its intrinsic link to SDPs. Specifically, a polynomial $p$ of degree $2d$ is SOS if and only if $p(x) = z(x)^T Q z(x)$, where $Q \succeq 0$ and $z(x)$ is a vector of monomials up to order $d$. Thus, certifying that a polynomial is SOS reduces to the task of finding a psd matrix $Q$ subject to a finite set of linear equalities, thus taking the form in~\eqref{eq:sosopt-general}. Certificates of the form in~\eqref{sos_cert} will form the building block for our approach. \vspace{-1em}

\section{SOS-Based Robust Tracking}\label{sec:sos} \vspace{-.5em}
\subsection{Relative System}\label{subsec:Relsys} \vspace{-.5em}
Given the tracking~ and planning models as in equation \eqref{tracker}, we define the {\em relative system state} $\rstate \in \reals^{n_r}$ as $\rstate := \rtrans(\tstate, \pstate)(\tstate - \ptmat\pstate)$, where $\rtrans(\cdot,\cdot): \reals^{n_s} \times \reals^{n_p} \rightarrow \reals^{n_r \times n_s}$, with $n_r \leq n_s$, is a matrix-valued map with bounded derivatives, and $\ptmat = \left[ \eye_{n_p}, \zero_{n_p\times (n_s-n_p)}\right]^\top$ is a projection matrix from $\reals^{n_p}$ to $\reals^{n_s}$ ($\eye$ and $\zero$ denote the identity and zero matrices, respectively). 
In the definition of $\ptmat$, we leveraged the assumption that the planning state $\pstate$ is a strict subset of the tracking state $\tstate$.  We make the following assumption on the dynamics of the relative system state, which is central for our approach:

\begin{assumption}[Relative System Dynamics]\label{assum:relative}
The relative system dynamics can be written as a Lipschitz continuous function $\rdyn$ of $(r, \tctrl, \pctrl)$, that is, $\dot\rstate = \rdyn(r,\tctrl,\pctrl)$.
\end{assumption}
\revision{Lipschitz continuity is necessary for the existence and uniqueness of solutions to the ODEs. The structural property in} Assumption \ref{assum:relative} is satisfied for a number of dynamical models of mobile robotic systems. 
For example, for ubiquitous cases where the planning model has integrator dynamics, Assumption \ref{assum:relative} is satisfied by simply selecting $\rtrans(\cdot,\cdot)$ as the identity map. However, in general, it is difficult to characterize the conditions on the planning/tracking  models such that Assumption \ref{assum:relative} is satisfied.
Therefore, in Table~\ref{tab:cat} in Appendix~\ref{sec:table}, we provide  a ``catalog" of tracking and planning models fulfilling Assumption \ref{assum:relative} for a representative set of robotic systems, along with the map $\rtrans$ that guarantees fulfillment of Assumption \ref{assum:relative} and the resulting relative system dynamics. Specifically, the relative system states in the 6th column are obtained by combining the tracking state in the 2nd column, planning state in the 4th column, and the transformation in the 5th column via $\rstate = \rtrans(\tstate, \pstate) (\tstate - \ptmat \pstate)$. Henceforth, we will consider planning/tracking models that fulfill Assumption \ref{assum:relative}, and refer to the system $\dot\rstate = \rdyn(\rstate,\tctrl,\pctrl)$ as the {\em relative system}. 

Finally, we define the {\em error state} $\estate$ as the relative system state \textit{excluding} the absolute states of the tracking model, and the {\em auxiliary state} $\astate$ as the relative system state \textit{excluding} the error state.
Hence, $\rstate = (\estate, \astate)$.
For instance, for the case of 
a 5D car model used to track a 3D Dubins car model  (second example in Table~\ref{tab:cat}), $\estate = (x_r, \, y_r, \,\theta_r)$ and $\astate=(v, \,\omega)$. Roughly speaking, considering relative system dynamics allows one to reformulate the tracking problem as a stabilization problem to the origin. \vspace{-1.5em}

\subsection{Optimization Problem} \vspace{-.5em}
In order to define a notion of \emph{tracking error bound} (TEB), let $c(\rstate): \reals^{n_r} \mapsto  \reals^{n_{\rt}}$ be a smooth mapping representing a vector of quantities we wish to remain bounded, with $n_{\rt} \geq n_r$; henceforth, we will refer to $c(\rstate)$ as the \emph{bounded state}. The simplest example for $c(\cdot)$ is the identity map. However, for more complex planning/tracking models, the mapping $c(\cdot)$ can be more sophisticated \revision{to better capture the structure of the planner/tracker dynamics pair} -- some examples are provided in Table~\ref{tab:cat}. 
The goal then is to find a closed and bounded set $\teb$ in the bounded state such that $c(\rstate(t)) \in \teb$ for all $t\geq 0$ (as an illustration, see the light blue ball in Figure~\ref{fig:intropic}). Such an invariance condition is captured by the  implication \vspace{-.4em}
\begin{equation}\label{eq:invar}
	c(\rstate(0)) \in \teb \Rightarrow c(\rstate(t)) \in \teb, \quad \forall t \geq 0. \vspace{-.4em}
\end{equation}
To parameterize this set, let $\rho$ be a positive constant and $\vft(\rt): \reals^{n_{\rt}} \mapsto  \reals$ be a smooth function with a bounded $\rho-$sublevel set. Also, define the function $\vf(\rstate ): \reals^{n_r} \mapsto \reals$ as simply the composition $\vft(c(\cdot))$. Suppose that the following implication is true: \vspace{-.5em}
\begin{align*}\small
\vf(\rstate) = \rho \quad \Rightarrow \quad \dot\vf =  \left.\frac{\partial\vft(\rt)}{\partial\rt}\right|^T_{\rt = c(\rstate)} \frac{\partial c(r)}{\partial \rstate} \rdyn(\rstate, \tctrl, \pctrl) < 0. \vspace{-.7em}
\end{align*}   
That is, there exists some tracking control $\tctrl$ such that the time-derivative of $V$ on the boundary $\vf(\rstate) = \rho$ is negative. It follows that the set $\{\rstate : \vf(\rstate) \leq \rho \}$ is invariant. Then, the set $\{ c(\rstate) : \vft(c(\rstate)) \leq \rho\}$ is a valid error bound $\teb$ in the sense of equation~\eqref{eq:invar}. An illustration of this idea is provided in Fig.~\ref{fig:intropic} (right). 

\begin{remark}
Note that the set $\{c(\rstate) : \vft(c(\rstate)) \leq \rho\}$ is equivalent, up to a translation and scaling factor, to the set  $\{c(\rstate) : \alpha \left( \vft(c(\rstate))  )+ \beta\right) \leq \alpha (\rho + \beta) \}$, where $\beta \in \reals$ and $\alpha > 0$.  To eliminate this redundancy, we impose, without loss of generality, the conditions $\vft(0) = 0$ and $\rho = 1$. The choice $\vft(0) = 0$  essentially corresponds to a  translational adjustment that ``centers'' the set $\teb$ on a zero bounded state. 
\end{remark}

In the context of this paper, set $\teb$ is used as a ``buffer" that a planner should consider along each candidate nominal motion plan, to ensure that the {\em realized} tracking trajectory is safe, that is, collision free. Clearly, the search for set $\teb$ is intertwined with the search for a tracking controller that can keep the realized tracking trajectory within it. We parameterize the tracking controller $\tctrl$ as a function of the relative system state and planning control signal, that is $\tctrl = \fbctrl(\rstate, \pctrl)$. 
Denote the closed-loop dynamics as $\rdyn_\fbctrl(\rstate, \pctrl) := \rdyn(\rstate, \fbctrl(\rstate, \pctrl), \pctrl)$ and define
\[
	\dot{V}_K(\rstate,\pctrl) :=  \left.\frac{\partial\vft(\rt)}{\partial\rt}\right|^T_{\rt = c(\rstate)} \frac{\partial c(\rstate)}{\partial \rstate} \rdyn(\rstate, \fbctrl(\rstate, \pctrl), \pctrl). 
\]
The search for a suitable function  $\vf$ and controller $K(\cdot,\cdot)$ is then formalized by the optimization problem: \vspace{-1em}
\begin{subequations}\label{eq:opt0}  \small
  \begin{align}
  \underset{\fbctrl(\cdot,\cdot),\teb }{\text{minimize}}\quad &\text{volume}(\teb) \label{eq:opt0:obj} \\
  \text{subject to}\quad & 
  \begin{aligned}[t]
  \begin{rcases}
  &\valfunc(\rstate) = 1 \\
	&\pctrl \in \pcset 
	\end{rcases} \Rightarrow \dot \vf_\fbctrl(\rstate, \pctrl) < 0 
  \end{aligned} \label{eq:opt0:lya} \\
  \quad &
   \begin{aligned}[t]
   \begin{rcases}
  &\valfunc(\rstate) \leq 1 \\
	&\pctrl \in \pcset 
	\end{rcases} \Rightarrow \fbctrl(\rstate, \pctrl)\in \tcset 
  \end{aligned} \label{eq:opt0:ctrl}
  \end{align}
\end{subequations}

We will next encode the constraints and objective of problem  \eqref{eq:opt0} as SOS certificates.\vspace{-1.2em}

\subsection{Reformulating the Constraints as SOS Certificates}\label{subsec:SOSconstraints}
\vspace{-.2em}
We consider the constraints first. Under the assumption that the dynamics for the tracking model are control affine \revision{(common for most robotic systems)}, the function $\rdyn$ is control affine in $\tctrl$. That is, $\rdyn$ takes the form $\rdyn(\rstate,\pctrl,\tctrl) = h(\rstate,\pctrl) + B(\rstate) \tctrl$.
Thus, \vspace{-0.5em}
\[
	\dot V = \frac{\partial\valfunc(\rstate)}{\partial\rstate}^T h(\rstate,\pctrl) + \frac{\partial\valfunc(\rstate)}{\partial\rstate}^T B(\rstate) \tctrl. \vspace{-.5em}
\]
Hence, the invariance condition requires the existence of a tracking controller $\fbctrl(\rstate,\pctrl)$ such that \vspace{-0.4em}
\begin{equation}
	\begin{rcases}
	&\valfunc(\rstate) = 1 \\
	&\pctrl \in \pcset 
	\end{rcases} \Rightarrow \frac{\partial\valfunc(\rstate)}{\partial\rstate} ^T h(\rstate,\pctrl)  + \frac{\partial\valfunc(\rstate)}{\partial\rstate}^T B(\rstate) K(\rstate,\pctrl) < 0. \vspace{-.3em}
\label{new_eq_1}
\end{equation}
We can \emph{equivalently} state inequality \eqref{new_eq_1} as:
\begin{equation}  \small
	\begin{rcases}
	&\valfunc(\rstate) = 1 \\
	&\pctrl \in \pcset 
	\end{rcases} \Rightarrow   \frac{\partial\valfunc(\rstate)}{\partial\rstate}^T h(\rstate,\pctrl) + \min_{\fbctrl(\rstate,\pctrl) \in \tcset}   \left( B(r) ^T \, \frac{\partial\valfunc(\rstate)}{\partial\rstate} \right) ^T  \fbctrl(\rstate,\pctrl) < 0. \vspace{0em}
\label{new_eq_2}
\end{equation}
The equivalence between inequalities \eqref{new_eq_1} and \eqref{new_eq_2} follows from the observation that inequality~\eqref{new_eq_1} holds for \emph{some} $\tctrl \in \tcset$ if and only if it holds for a $\tctrl \in \tcset$ minimizing the left hand side in inequality \eqref{new_eq_2}.  Importantly, the minimization in inequality  \eqref{new_eq_2}  is \emph{independent} of $\pctrl$, as the constraint set $\tcset$ and objective coefficient vector $B^T \partial \valfunc/\partial \rstate$ do not depend on $\pctrl$. Thus, we can simplify $\fbctrl$ by making it a function of $\rstate$ only (with a slight abuse of notation, we still refer to such a simplified tracking controller with $\fbctrl$). 

We can now 
define SOS certificates for the constraints in problem~\eqref{eq:opt0}. Suppose set $\pcset$ can be written as the semialgebraic set $\{\pctrl \in \reals^{m_p} : \setfunc_i^\pstate(\pctrl) \le 0, \ i = 1,\ldots, N_p\}$ and let the controller $\fbctrl$ be a \emph{polynomial} function in $\rstate$. Expanding $\dot\vf_\fbctrl$, one can encode constraint~\eqref{eq:opt0:lya} using the multiplier polynomials $\ \laglya(\rstate, \pctrl), \{\lagpc_i (\rstate,\pctrl)\}_{i=1}^{N_p}$ such that \vspace{-.8em}
\begin{align} \small \label{eq:Vdot_constr}
-\dot{\vf}_K  + \laglya \cdot (\valfunc-1) + \sum_{i=1}^{N_p} \lagpc_i \cdot \setfunc_i^\pstate \issos, &&
\{\lagpc_i\} \text{ are SOS}, && i = 1,\ldots,N_p.  \vspace{-1.5em}
\end{align}
(In the following we drop the summation notation and write the collective sum as a ``dot-product:''  $\lagpc \cdot \setfunc^\pstate $). To capture the control constraint in \eqref{eq:opt0:ctrl}, one may leverage two different techniques. The first one is to compose the tracking controller $\fbctrl$ with a saturation function and write $\dot{\vf}$ in case form corresponding to the unsaturated and saturated regimes, respectively. However, this approach scales exponentially in the dimension of the control input, as one needs to capture all combinations of saturation regimes \cite{MajumdarTedrake2017}. Instead, we assume that the tracking control set $\tcset$ is polytopic, i.e., $\tcset = \{ \tctrl \in \reals^{m_s}:  \setfunc^\tstate(\tctrl)\le0\}$, where $\setfunc^\tstate(\tctrl)\le0$ is a set of linear inequalities in $\tctrl$ of the form $\setfunc^\tstate_i = a_{s_i}^T \tctrl - b_{s_i} \leq 0$, $i = 1,\ldots,N_s$. Note that this is not an overly restrictive assumption as the control constraints are often taken to be box constraints. Then, we encode constraint~\eqref{eq:opt0:ctrl} using the multipliers $\{\lagtc_i(\rstate, \pctrl)\}$:
\begin{align} \label{eq:ctrl_constr}
-\setfunc_i^\tstate(\fbctrl)  + \lagtc_i \cdot (\valfunc-1) \issos, &&
\lagtc_i \text{ are SOS}, && i = 1,\ldots, N_s.
\end{align}
Crucially, this approach scales linearly with respect to the number of inputs. \vspace{-1em}

\subsection{Reformulating the Objective as SOS Certificates} \label{sec:obj_SOS} \vspace{-.3em}
Minimizing the volume of $\teb$, which itself is a non-linear mapping of the $1-$sublevel set of $\vf$, is a difficult task if one reasons in terms of arbitrary functions $\vf$. One approach is to approximate the volume by the integral of $\vf$ over an Euclidean ball of radius $R$ such that the ball is contained within the $1-$sublevel set of $\vf$ -- an approach taken, for example, in~\cite{PosaKoolenEtAl2017}. Alternatively, one can minimize the volume of an encapsulating ellipsoid, easily captured using an additional constraint and a convex pseudo-objective. 

Toward this end, we define the ellipsoid  $\elps := \{\rstate: c(\rstate)^\top \elpsf c(\rstate) \le 1\}$ for some positive definite matrix $\elpsf \succeq \delta_E I$, where $\delta_E > 0$ is a small tolerance parameter. Then, the inclusion constraint $\teb \subseteq \elps$ is captured using the SOS multiplier $\lagelps(\rstate)$: \vspace{-.5em}
\begin{align}  \small
	1 - c^\top \elpsf c + \lagelps(\vf-1) \issos, &&
	\lagelps \issos, &&
	\elpsf \succeq \delta_E I.
\label{eq:V_ell_r}
\end{align}
As the volume of $\elps$ is inversely proportional to the square root of its determinant, the minimum volume objective reduces to maximizing the determinant of $E$, and can be written in the form of~\eqref{eq:sosopt-general}~\cite[Chapter 4]{Ben-TalNemirovski2001}. For non-identity mappings $c(\cdot)$, the first SOS constraint in equation \eqref{eq:V_ell_r} can quickly become computationally challenging due to the quadratic terms in $c(r)$ and the complexity of the relation $V(\cdot) = \vft (c (\cdot))$. In this case, one may consider the following simplification. Since\vspace{-.5em}

\begin{equation}
	\teb = \{ c(\rstate) \in \reals^{n_{\rt}}: \vft (c(\rstate)) \leq 1 \} \subseteq \{ \rt \in \reals^{n_{\rt}}: \vft (\rt) \leq 1 \},\vspace{-.2em}
\label{teb_incl}
\end{equation}
the simplification is to outer bound the set on the right by using an ellipsoid. Specifically, let the ellipsoid $\elps$ be given by $\elps = \{\rt: \rt^\top \elpsf \rt \le 1\}$ for some positive definite matrix $\elpsf \succeq \delta_E I$. The inclusion condition $\vft(\rt) \le 1 \Rightarrow \rt^\top \elpsf\rt \le 1$, is captured using the SOS multiplier $\lagelps(\rt)$: \vspace{-0.5em}
\begin{align}  \small
	1 - \rt^\top \elpsf \rt + \lagelps(\vft-1) \issos, &&
	\lagelps \issos, &&
	\elpsf \succeq \delta_E I.
\label{eq:V_ell}
\end{align}
Crucially, the constraints above are deliberately written with respect to the variable $\rt$, which is treated as an {\em independent indeterminate} from $\rstate$. This is beneficial when using a mapping $c(\cdot)$ other than the trivial identity map, as it allows one to approximate the otherwise complex set $\teb$ with potentially simpler functions in $\rt$. \vspace{-1em} 

\subsection{SOS Formulation of Optimization Problem}

Collecting all the results so far, our approach entails conservatively (as we rely on sufficient SOS certificates) solving the optimization problem \eqref{eq:opt0} as a SOS program:\vspace{-.5em}
\begin{subequations}\label{eq:opt2}  \small
  \begin{align}
  \underset{\fbctrl, \vft, \elpsf, \mathbf\lag}{\text{maximize}}\quad & \log\det(\elpsf) \label{eq:opt2:obj} \\
  \text{subject to}\quad & \text{eqs.~\eqref{eq:Vdot_constr},~\eqref{eq:ctrl_constr},~\eqref{eq:V_ell}} \label{eq:opt2:cons} \\
  & \vft(0) =0,
  \label{eq:non_empty}
  \end{align}
\end{subequations}
\noindent where $\mathbf\lag := \{ \laglya, \lagpc, \lagtc, \lagelps\}$. Reinforcing the ideas in Section~\ref{sec:obj_SOS}, we iterate that $\vft$ is considered as a function in the independent indeterminate $\rt$, while constraints~\eqref{eq:Vdot_constr},~\eqref{eq:ctrl_constr} are encoded using the indeterminate $\rstate$ via the definition $V(\rstate) = \vft(c(\rstate))$. 

Constraints in~\eqref{eq:opt2:cons} are bilinear in the decision variables. Consequently, similar to the bilinear solution algorithms in~\cite{MajumdarTedrake2017} and~\cite{PosaKoolenEtAl2017}, one must alternate between the decision variable sets $\{E, K, \mathbf\lag \}$ and $\{\vft, E, \lagpc\}$, each time holding the other variable set fixed. Specifically, we refer to the sub-problem where $K(\cdot)$ is part of the decision variable set as the $K$ sub-problem, and the sub-problem where $\vft$ is part of the decision variable set as the $V$ sub-problem.  Direct implementation of alternations is hindered by two challenges. First, one requires a feasible initial guess for $\vft$ to begin the alternations. In \cite{MajumdarTedrake2017} and~\cite{PosaKoolenEtAl2017}, the authors leverage locally linearized models and the solution to the Riccati equation to generate such a guess. In this paper we address a fundamentally more challenging problem as we consider the controllability of a relative dynamical system between two different dynamical models. Consequently, the relative system is often not linearly controllable at $r=0$ (even if the individual models are) and thus one requires an additional procedure to {generate} a feasible function $\vft$ for problem~\eqref{eq:opt2} from an initially infeasible guess. Second, alternating optimization is prone to numerical instabilities as the solutions of the individual convex problems frequently lie on boundaries of the respective feasible sets. We address these challenges next. \vspace{-1em}

\section{Solving the Bilinear Optimization Problem}\label{sec:solve_sos}\vspace{-1em}

The general idea to tackle both of the aforementioned challenges entails using iterative slack minimization. We first discuss the solutions to the $K$ and $V$ sub-problems (Sections \ref{subsec:Kprob} and \ref{subsec:Vprob}, respectively) and then present the general solution algorithm in Section \ref{subsec:SolAlgo}.\vspace{-1em}


\subsection{The $K$ Sub-Problem}\label{subsec:Kprob}\vspace{-.5em}

A na\"ive implementation of the $K$ sub-problem would entail solving problem~\eqref{eq:opt2} with respect to $\{E, K, \mathbf{L}\}$ for a given $\vft$. However, constraint~\eqref{eq:Vdot_constr} as written may generate controllers that are numerically unstable when held fixed in the $V$ sub-problem. Given the polytopic constraint set $\tcset$ and a fixed $\vft$, one can exactly characterize the ``most stabilizing controller'' as the function $K$ that minimizes the left hand side in inequality~\eqref{new_eq_2} for all $\rstate$ such that $V(\rstate) = 1$. Thus, we propose the following two-step method for solving the $K$ sub-problem, \emph{for a given function} $\vft$:\vspace{-.5em} 
\begin{enumerate}

\item Find the ``most stabilizing controller'' $K(\cdot)$ as the solution to:\vspace{-.5em} 
\begin{subequations}\label{eq:opt_sub_1a}
  \small
  \begin{align}
  \underset{K, \laglya, \lagpc, \lagtc, \gamma}{\text{minimize}}\quad & \gamma   \\
  \text{subject to}\quad & - \dot{\vf}_K + \gamma + \laglya \cdot (\vf - 1) + \lagpc \cdot \setfunc^\pstate \quad \issos \\
  & - \setfunc^\tstate(\fbctrl)  + \lagtc \cdot (\vf-1) \issos,\qquad \{\lagpc_i\}, \{\lagtc_i\} \quad \text{are SOS},
  \end{align}
\end{subequations}
\noindent where $\gamma$ is a slack variable for the invariance constraint. Notice that when $\vf(\rstate) = 1$ and $\setfunc^\pstate (\pctrl) \leq 0$, then $\dot{\vf}_K (\rstate,\pctrl) \leq \gamma$. Denote the optimal slack as $\gamma^*_c$.

\item Compute the tightest bounding ellipsoid $\elps$:\vspace{-.5em} 
\begin{subequations}\label{eq:opt_sub_1c}
    \small
  \begin{align}
  \underset{\lagelps, E}{\text{maximize}}\quad & \log\det(\elpsf)   \\
  \text{subject to}\quad & 1 - \rt^\top \elpsf \rt + \lagelps(\vft-1) \quad \issos \\
	&\elpsf \succeq \delta_E I, \qquad \lagelps \issos.
  \end{align}
\end{subequations}

\end{enumerate}

The two steps above comprise the $K$ sub-problem. The benefit of decomposing the $K$ sub-problem in this fashion is that we independently search for the most stabilizing controller while simultaneously relaxing the invariance constraint by using the slack variable $\gamma$; in particular, $\gamma$ accounts for the suboptimality of the computed controller with respect to the left hand side of inequality~\eqref{eq:Vdot_constr}.\vspace{-1em}

\subsection{The $V$ Sub-Problem}\label{subsec:Vprob}
Given a controller $K(\cdot)$, and multiplier polynomials $\{\lagtc_i\}$, $\lagelps$, and $\laglya$ from the $K$ sub-problem, the $V$ sub-problem is defined as\pagebreak
\begin{subequations}\label{eq:opt_sub_2}
  \small
  \begin{align}
  \underset{\vft, E, \lagpc, \gamma, \epsilon }{\text{maximize}}\quad & \lambda \log\det(\elpsf) -  \gamma - \|\epsilon\|_1   \\
  \text{subject to}\quad & - \dot{\vf}_K + \gamma + \laglya \cdot (\vf - 1)  + \lagpc \cdot \setfunc^\pstate \quad \issos  \label{opt_2_c1}  \\
    	& - \setfunc^\tstate(\fbctrl) + \epsilon + \lagtc \cdot (\vf-1) \issos \\
  	&1 - \rt^\top \elpsf \rt + \lagelps(\vft-1) \quad \issos \\
	&\elpsf \succeq \delta_E I, \quad \vft(0)=0, \\
  	& \{\lagpc_i\} \quad \text{are SOS} \label{opt_2_cf}
  \end{align}
\end{subequations}

\noindent where $\epsilon \in \reals^{N_s}_{\geq 0}$ is a slack vector for the control constraints and $\lambda \in \reals_{>0}$ is a Pareto trade-off parameter. Notice that the control slack $\epsilon$ is only necessary in the $V$ sub-problem since the controller is held fixed within this problem. In contrast, in problem~\eqref{eq:opt_sub_1a}, we directly optimize over the set of strictly \emph{feasible} controllers. Given these slack-based relaxed sub-problems, we now provide a solution algorithm for problem~\eqref{eq:opt2}. \vspace{-1em}


\subsection{Solution Algorithm}\label{subsec:SolAlgo}\vspace{-.5em} 

Before providing the full solution algorithm, we describe an initialization procedure that generates a numerically stable initial guess for $\vft$. The procedure is detailed  in Algorithm~\ref{alg:infeas}. In particular, the algorithm may be initialized by using any polynomial guess for $\vft$ such that $\vft(0) = 0$ (e.g., $\rt^T \rt$).
\begin{algorithm}[h!]
  \caption{Generating Feasible Guess}
  {\small
  \label{alg:infeas}
  \begin{algorithmic}[1]
  \State {\bf Input:} Initial guess $\vft$ satisfying $\vft(0) = 0$; slack tolerance $\delta \in [0,1)$; max \# of iterations $N$.
   \State $i \leftarrow 0$.
   \While {$i<N$} 
   \State $\{K, \laglya, \lagtc, \lagelps, E, \gamma_c^*\} \leftarrow $ \textproc{Solve} ~\eqref{eq:opt_sub_1a}--\eqref{eq:opt_sub_1c}. \label{subprob_1}
   \IfThen{$\gamma_c^* \leq \delta$} {\textbf{return} ($\vft,E, K$)} 
   \State $\{\vft, E, \gamma^*, \epsilon^* \} \leftarrow $ \textproc{Solve} ~\eqref{eq:opt_sub_2} with $\lambda = 0$ and subject to additional constraint $\gamma \leq \gamma_c^*$. \label{line:infeas_2}
  \IfThen{$\max\{\gamma^*, \|\epsilon^*\|_{\infty} \} \leq \delta$} {\textbf{return} ($\vft,E,K$)} 
   \State $i \leftarrow i+1$.
  \EndWhile
  \State  \Return Failure.
      \end{algorithmic}} 
\end{algorithm}  \vspace{-1em}

Notice that in line~\ref{line:infeas_2}, when problem~\eqref{eq:opt_sub_2} is solved, the Pareto parameter $\lambda$ is set to 0 since the objective for the initialization algorithm is to simply generate a feasible initial guess for problem~\eqref{eq:opt2}. Furthermore, we impose the additional constraint $\gamma \leq \gamma_c^*$ to ensure monotonic improvement in solution quality between sub-problems. As the original problem is still non-convex, the convergence of the slack variables  below the tolerance threshold is not guaranteed for every initial guess of $\vft$. However, typical guesses such as $\rt^T Q \rt$ for a positive diagonal matrix $Q$ appear to work quite well, as the experiment section will  illustrate. 

Given an initial guess generated by Algorithm~\ref{alg:infeas}, we are now ready to solve problem~\eqref{eq:opt2}. To do so, we will make use of a slightly modified version of the $V$ sub-problem, given below:  \vspace{-0.5em}
\begin{subequations}\label{eq:opt_sub_2_2} \small
  \begin{align}
  \underset{\vft, E, \lagpc, \gamma, \epsilon}{\text{maximize}}\quad &  \log\det(\elpsf) -   \lambda(\gamma + \|\epsilon\|_1)  \\
  \text{subject to}\quad & \text{eqs.~\eqref{opt_2_c1} --~\eqref{opt_2_cf}} \\
  & \underline{\alpha} E^* \preceq E \preceq (1+\alpha) E^*, \label{trust_region}
  \end{align}
\end{subequations}
where $E^*$ is the previous numerically stable solution (i.e., with optimal slack values $\gamma^*, \|\epsilon^*\|_{\infty} \leq \delta$), $\underline{\alpha}\in(0,1)$ is a fixed parameter, and $\alpha\in(0,1)$ is a backtracking search parameter that is adjusted in an iterative fashion. Note that the Pareto parameter $\lambda$ now multiplies the slack terms in the objective. Specifically, we iteratively solve problem~\eqref{eq:opt_sub_2_2}, \emph{each time} checking the slack tolerances to ensure numerical stability, while using constraint~\eqref{trust_region} to enforce a \emph{trust region} around the current numerically stable solution. The full solution algorithm is summarized in Algorithm~\ref{alg:feas}.\vspace{-1.5em}
\begin{algorithm}[h!]
  \caption{Solving Problem~\eqref{eq:opt2}}
  \label{alg:feas}
 {\small
  \begin{algorithmic}[1]
  \State {\bf Input:} Output tuple $(\vft, E, K)$ from Algorithm~\ref{alg:infeas}; slack tolerance $\delta$; termination tolerance $\theta_1 \in(0,1)$.
   \State $i\leftarrow 1$, \texttt{converged} $\leftarrow$ \texttt{false}.
   \State \textbf{Initialize}: $(\vft^*, E^*, K^*, c^*_0) \leftarrow (\vft,E,K, \log\det(E))$.
   \While {$i < N \wedge \neg$ \texttt{converged}} 
   \State $\{K, \laglya, \lagtc, \lagelps, E\} \leftarrow $ \textproc{Solve}~\eqref{eq:opt_sub_1a}--~\eqref{eq:opt_sub_1c} subject to additional constraint: $\gamma \leq 0$. \label{line:subprob_1} 
   \If {Line~\ref{line:subprob_1} successfully solved} 
	\State $(E^*, K^*) \leftarrow (E,K) $.
	\State $(\vft^*, E^*,  c_i^*) \leftarrow$ \texttt{Backtrack}$\left(\{K^*,\laglya,\lagtc, \lagelps\},E^*, \vft^*\right)$. \label{line:back}
	\IfThenElse {$|c^*_i - c^*_{i-1}| \leq \theta_1 |c^*_{i-1}|$} 
	{\texttt{converged} $\leftarrow$ \texttt{true}.}
	{$ i \leftarrow i+1$.}
  \Else
  	\State \texttt{converged} $\leftarrow$ \texttt{true}.
  \EndIf
  \EndWhile
  \State \Return $(\vft^*,  E^*, K^*)$
   \end{algorithmic}
      }
\end{algorithm}\vspace{-3.5em}

\begin{algorithm}[h!]
  \caption{Backtrack}
  \label{alg:back}
 {\small
  \begin{algorithmic}[1]
  \State {\bf Input:} Solution $\{K^*,\laglya,\lagtc, \lagelps\}$ and $E^*$ from line~\ref{line:subprob_1} in Algorithm~\ref{alg:feas} and current best solution $\vft^*$; slack tolerance $\delta$; backtrack parameters $\overline{\alpha}, \, \beta,\, \theta_2\in(0,1)$.
   \State \textbf{Initialize}: $\alpha \leftarrow \overline{\alpha}$, \texttt{bt}$\leftarrow$ \texttt{false}, $c^* \leftarrow \log\det(E^*)$.
   \While {$\neg$\texttt{bt}}  \label{line:bt_1_s}
   \State $\{\vft, E, \gamma^*, \epsilon^* \}  \leftarrow $ \textproc{Solve} ~\eqref{eq:opt_sub_2_2}. 
	 \IfThenElse{$\max\{\gamma^*, \|\epsilon^*\|_{\infty} \} \leq \delta $}
		 {$(\vft^*, E^*,  c^*) \leftarrow (\vft,E,\log\det(E))$.}
   		{\texttt{bt}$\leftarrow$ \texttt{true}.}
  \EndWhile \label{line:bt_1_e}
    \While {\texttt{bt} $\wedge \ \alpha>\theta_2 \overline{\alpha}$} \label{line:bt_2_s}
    \State $\alpha \leftarrow \beta \alpha$.
   \State $\{\vft, E, \gamma^*, \epsilon^* \} \leftarrow $ \textproc{Solve} ~\eqref{eq:opt_sub_2_2}.
	 \If{$\max\{\gamma^*, \|\epsilon^*\|_{\infty} \} \leq \delta $}
	 	\State $ (\vft^*, E^*,  c^*) \leftarrow (\vft,E,\log\det(E))$.
	 	\State \texttt{bt} $\leftarrow $\texttt{false}.
	\EndIf
  \EndWhile   \label{line:bt_2_e}
  \State \Return $(\vft^*, E^*,  c^*)$
      \end{algorithmic}
      }
\end{algorithm}\vspace{-2em}

The backtrack search procedure in line~\ref{line:back} is summarized  in Algorithm~\ref{alg:back}. Within this procedure, we first iteratively maximize $\log\det(E)$ within the trust region~\eqref{trust_region} centered on the previous stable numerical solution, using the slack values as a check on solution quality (lines~\ref{line:bt_1_s}---\ref{line:bt_1_e}). Once solution quality degrades, we backtrack (shrink) the trust region in lines~\ref{line:bt_2_s}---\ref{line:bt_2_e} until we again fall below the slack tolerances. Note that Algorithm~\ref{alg:back} will either return an updated $\{\vft,E\}$ that is numerically stable (with respect to slack tolerances), or it will simply return the numerically stable solution from line~\ref{line:subprob_1} in Algorithm~\ref{alg:feas} if unable to make progress. Thus, Algorithm~\ref{alg:feas} terminates if either (1) improvement in $\log\det(E)$ stalls, or (2) the function $\vft^*$ returned by the backtrack procedure is not \emph{strictly} feasible (i.e., $\gamma \leq 0$) with respect to line~\ref{line:subprob_1} in Algorithm~\ref{alg:feas}, but is still acceptable according to the slack tolerances. 

The key difference between the alternating method described here and a similar procedure in~\cite{PosaKoolenEtAl2017} is that the authors in~\cite{PosaKoolenEtAl2017} minimize the slack variable in both sub-problems and use binary search to iteratively refine an upper bound on the cost function within the $V$ sub-problem. On the other hand, our algorithm maintains numerical stability by using the slack variables \emph{solely as a check} on the allowable change in the solution, while taking advantage of minimizing the true objective (i.e., $\log\det(E)$) within \emph{both} sub-problems. This is especially useful in situations where the second phase of the algorithm struggles to make notable improvements on the objective (e.g., due to a smaller set of decision variables), thereby allowing the $K$ sub-problem to take over.\vspace{-1em} 

\section{Numerical examples \label{sec:num}}
\vspace{-1em}
In this section we numerically validate our proposed approach. Specifically, in Section \ref{subsec:HJ}, we compare our approach with the HJ-based approach in \cite{HerbertChenEtAl2017}, while in Section \ref{subsec:num_hi}, we study a high-dimensional system which is beyond the reach of HJ analysis. Additional numerical examples are provided in Appendix~\ref{sec:add}. The code is available at \url{https://github.com/StanfordASL/Model-Mismatch}.\vspace{-1em}

\subsection{Comparison with the HJ Method}\label{subsec:HJ}
\begin{figure}
	\centering
	\begin{subfigure}[t]{\textwidth}
		\centering
		\includegraphics[width=.85\textwidth,clip]{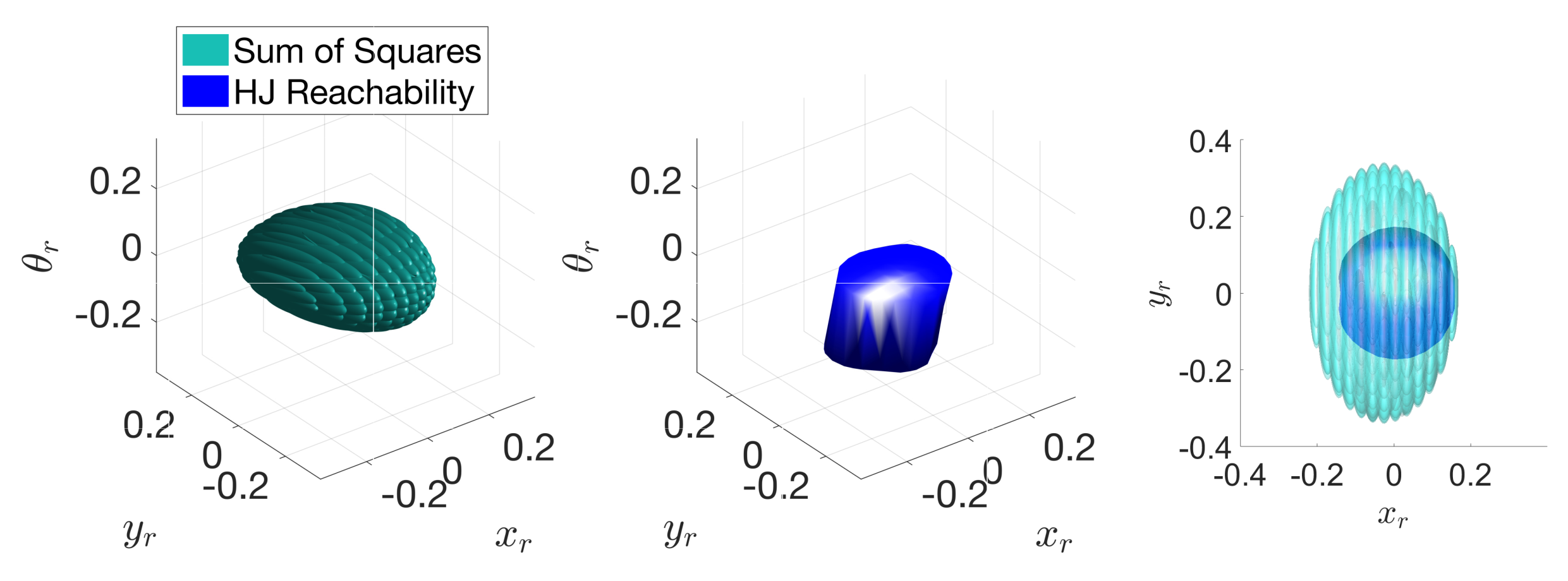}
		\label{fig:5D_V_pos}
	\end{subfigure}\hspace{0.1em}
	\caption{Projection of the boundary of $\teb$ onto $(x_r,y_r,\theta_r)$. {\em Left}: SOS. {\em Middle}: HJ. {\em Right}: Top-down view (i.e., onto $(x_r,y_r)$ for both solutions). The HJ positional error bound is smaller (42\% of SOS bound), but the SOS solution requires only 0.3\% of the computation time required by HJ analysis. 
	}
	\label{fig:5D_V}
	\vspace{-2em}
\end{figure}

\vspace{-.5em}
In this section, we compare our method with the HJ-based approach in~\cite{HerbertChenEtAl2017} for the case of a 5D car model used to track a 3D Dubins car model (that is, the second example in Table~\ref{tab:cat}). \revision{This example was chosen because five dimensions is the current limit for standard grid-based HJ reachability methods when techniques such as decomposition cannot be applied.} The system dynamics and bounded state definition are provided in the second row of Table 1. The model parameters were chosen as $|a| \leq 1$ m/s$^2$, $|\alpha|\le 3$ rad/s$^2$, $\hat v = 1.0$ m/s, $|\hat\omega|\le 0.1$ rad/s.
For the SOS method, we parameterized $K$ and $\vft$ as 2nd order polynomials in $\rstate$ and $\rt$, respectively. The trigonometric terms $\cos\theta_r$ and $\sin\theta_r$ were approximated with Chebyshev polynomials up to degree 2, over the range $\theta_r \in [-\pi/6,\pi/6]$ rad. To ensure the validity of these approximations, an additional constraint was appended to problems~\eqref{eq:opt_sub_1c} and~\eqref{eq:opt_sub_2}, namely: \vspace{-.5em}
\begin{align}
-\setfunc + L_g (\vft - 1) \issos, && L_g \issos,
\end{align}
\noindent where $g = \theta_r^2 - (\pi/6)^2$ and $L_g$ is a SOS polynomial in $\rt$. The initial guess for $\vft$ was simply $\rt^T Q \rt$, where $Q$ is a diagonal matrix with randomly sampled positive entries. In order to ensure a fair comparison, the cost function in the min-max game for the HJ method was $x_r^2 + y_r^2 + \theta_r^2 + (v\cos\theta_r - \hat v)^2 + v^2\sin^2\theta_r + \omega^2$, which is the closest analogue to the SOS objective of minimizing the entire volume of $\teb$ and not simply its projection onto the position coordinates. Figure~\ref{fig:5D_V} plots the projections of the boundary of set $\teb$ onto the $(x_r,\, y_r,\, \theta_r)$ components (i.e., the dimensions most relevant for collision checking) for the HJ and SOS solutions, respectively. The right-most panel in this figure provides a top-down view onto the $(x_r, \, y_r)$ plane. As expected, the HJ solution provides a tighter error bound -- approximately 42\% smaller than the SOS error bound, or about $0.2$ m in absolute terms. 
The main reason behind this is that by using a grid-based approach to solve the min-max HJ game, one \emph{exactly} computes the \emph{non-smooth} ``optimal controller'' in~\eqref{new_eq_2}, whereas the SOS approach attempts to find the best polynomial approximation. On the other hand, the computation time difference for the two solutions is \emph{substantial} -- approximately 25 hours for the HJ solution versus 5 minutes for the SOS solution. Specifically, the HJ solution was obtained by solving a corresponding HJ PDE \cite{HerbertChenEtAl2017} using a C++ implementation of the Lax Friedrichs numerical scheme \cite{TanabeChen2018}. The SOS programs were solved using the Spotless polynomial optimization toolbox~\cite{TobenkinPermenterEtAl2013} and MOSEK SDP solver~\cite{ApS2017}. 
Computations were done on a desktop computer with an Intel Core i7-2600K CPU and 16 GB of RAM. At least for this example, the SOS approach appears to provide a high-quality approximation to the optimal (HJ-based) solution, in a fraction of the time required by the HJ-based approach.

In Appendix~\ref{sec:add}, we provide an additional comparison between the HJ and SOS methods for a 4D dynamically extended Dubins car tracking a 2D single integrator planning model (row 1 in Table~\ref{tab:cat}). This example highlights a case in which the HJ method is efficient enough to warrant its use instead of the SOS method (i.e. when the dynamics are less than five dimensions and/or can be decomposed into subsystems that are each less than five dimensions). In this example, an optimal bound of 0.1 m between the two models was computed using HJ while SOS yielded a bound of 0.3 m. The computation time comparison was on the order of 2 minutes for SOS versus 5 minutes for HJ. The projection of the optimal $\teb$ onto the translational plane from the HJ method clearly illustrates the limitation of SOS in its ability to approximate non-smooth quantities. Indeed, herein lies the primary weakness of the SOS method -- tight approximations require high degree polynomials which can rapidly increase the size of the SOS problems. Despite this limitation inherent in the SOS method, the approach allows us to establish a conservative over-approximation of the tracking error bound (when one exists) for a given planner/tracker pair and gauge the sensitivity of the expected tracking error to changing model parameters such as inertial properties and control authorities. Furthermore, as the next section illustrates, the SOS method allows us to perform this analysis for higher dimensional systems that are severely challenging (albeit, not impossible) for HJ. As such, the HJ and SOS methods should really be viewed as complementary to one another.\vspace{-1em}

\subsection{High-Dimensional Example \label{subsec:num_hi}}
\vspace{-.3cm}
In this section, we present numerical results for a high-dimensional system for which the HJ approach is intractable. Specifically, we consider the case where an 8D plane model is used to track a 4D decoupled  Dubins plane model (that is, the fourth example in Table~\ref{tab:cat}).
\revision{To ensure that this example cannot be solved using standard grid-based HJ reachability, we must first determine that decomposition techniques would not be applicable. The 8D relative system dynamics are highly coupled, which means that approximate decomposition techniques would lead to results that are overly conservative \cite{ChenHerbertEtAl2018}. The dynamics also do not form self-contained subsystems, making exact decomposition impossible \cite{ChenHerbertEtAl2016b}.} 
This example well illustrates the usefulness of our approach. The planning model (i.e., decoupled Dubins plane) is a benchmark planning example with well characterized optimal trajectories~\cite{ChitsazLaValle2007}. The full 8D dynamics of the plane, however, are considerably more difficult to plan trajectories for online. Specifically, the system dynamics and bounded state definition are provided in the fourth row of Table~\ref{tab:cat}. The planner model parameters are as follows. The constant speed $\hat{v}$ is set to be the nominal speed for the plane in straight level flight conditions (lift equals gravitational force) at a nominal angle of attack $\alpha_0 = 5^o$, which corresponds to $\hat{v} = 6.5$ m/s (see~\cite{SchmerlingPavone2017} for the relevant constants used to compute this quantity). The maximum magnitude of the turning rate for the planner, $|\hat{\omega}|$, is set to be $20\%$ of the plane's maximum turning rate in a horizontal coordinated turn at nominal speed $\hat{v}$, and was computed to be $0.21$ rad/s (equivalently, a minimum turning radius of 30 m for the horizontal Dubins model). Finally, the planner's vertical velocity is limited to the range $[-0.1, \, 0.1]\,\hat{v}$. \vspace{-.15em}

Taking advantage of the structure of the dynamics, the normalized acceleration control $u_a$ was chosen to exactly cancel drag, plus an additional 2nd degree polynomial in $\rstate$ as determined by the SOS program. The rest of the controller components and $\vft$ were also parameterized as  2nd order polynomials in $\rstate$ and $\rt$, respectively. The plane's (tracking) control limits were chosen to be $u_a \in [-8,8]$ m/s$^2$ ($\approx 10$ times the acceleration needed to cancel drag at level trim conditions), $u_{\dot{\phi}} \in [-120,120]^o/$s, and $u_{\dot{\alpha}} \in [-60,60]^o/$s. To enforce the validity of the Chebyshev approximation for the trigonometric and $1/v$ terms in the dynamics, we enforce the additional constraints $\phi \in [-\pi/4, \,\pi/4]$ rad, $\gamma \in [-\pi/6,\,\pi/6]$ rad, $\psi_r \in [-\pi/6,\, \pi/6]$ rad, and $v \in [3,\, 10]$ m/s. The overall (initialization plus main optimization) computation time was under 2 hours. 
The projection of the set $\teb$ onto $(x_r, \, y_r, \, z_r)$ (the position errors \emph{in the frame of the Dubins car}) had span $[-5.6, \, 5.6] \times [-3.8, \, 3.8] \times [-4.5,\, 4.5]$ m. The bound is naturally more loose in the $(x_r,z_r)$ dimensions since the planning model's horizontal velocity is equal to the plane's trim conditions at level flight, limiting the plane's ability to use its remaining velocity range for both tracking and ascending/descending. Thus, for such a choice of planning and tracking models, the bound appears reasonable. To obtain tighter bounds, one could alternatively use the coupled kinematic plane model proposed in~\cite{OwenBeardEtAl2015} as the planning model. \vspace{-.15em}

\begin{figure}
	\centering
	\begin{subfigure}[t]{0.5\textwidth}
		\centering
		\includegraphics[width=\textwidth,clip]{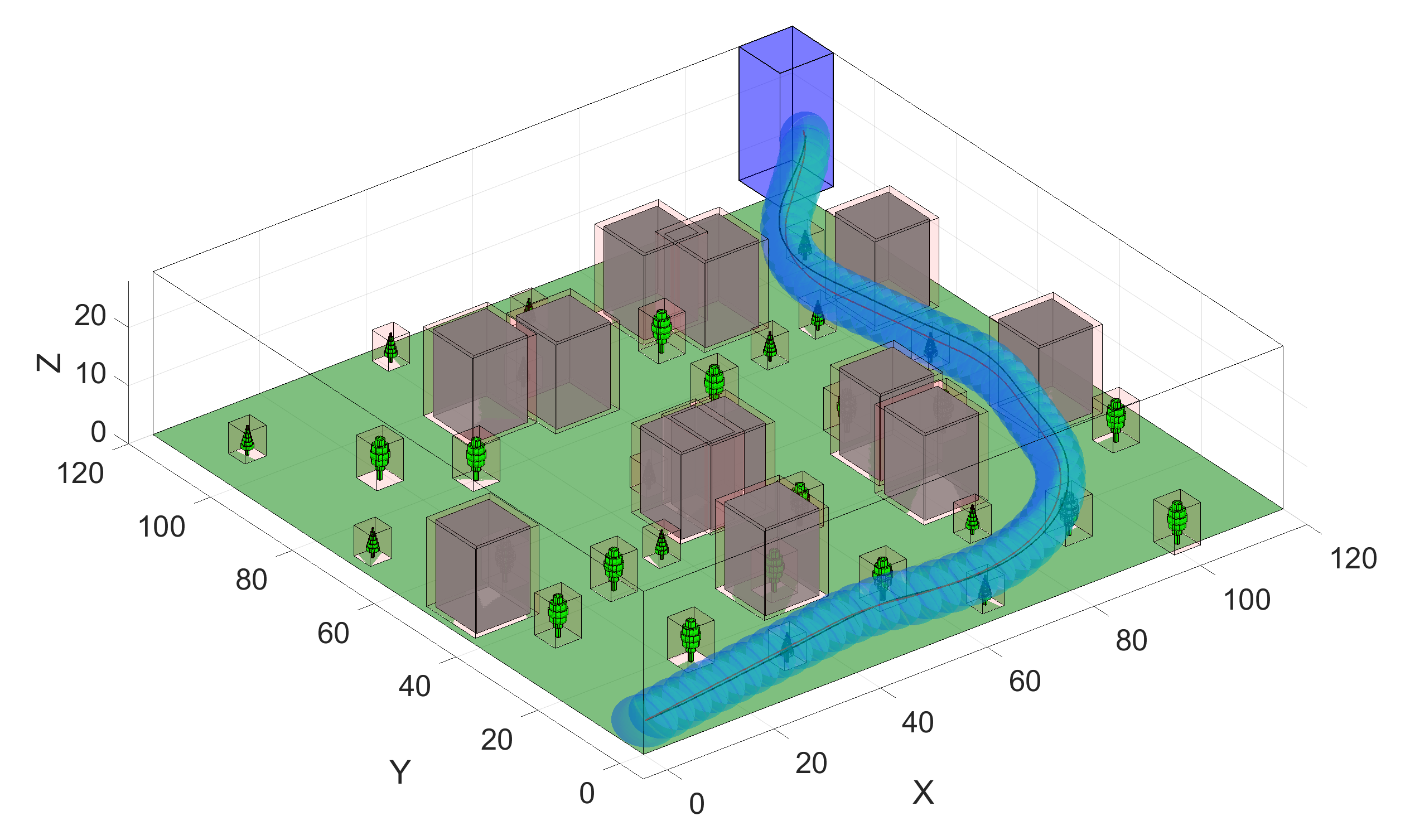}
		\caption{}
		\label{fig:plane_env}
	\end{subfigure} \qquad
	\begin{subfigure}[t]{0.44\textwidth}
		\centering
		\includegraphics[width=\textwidth,clip]{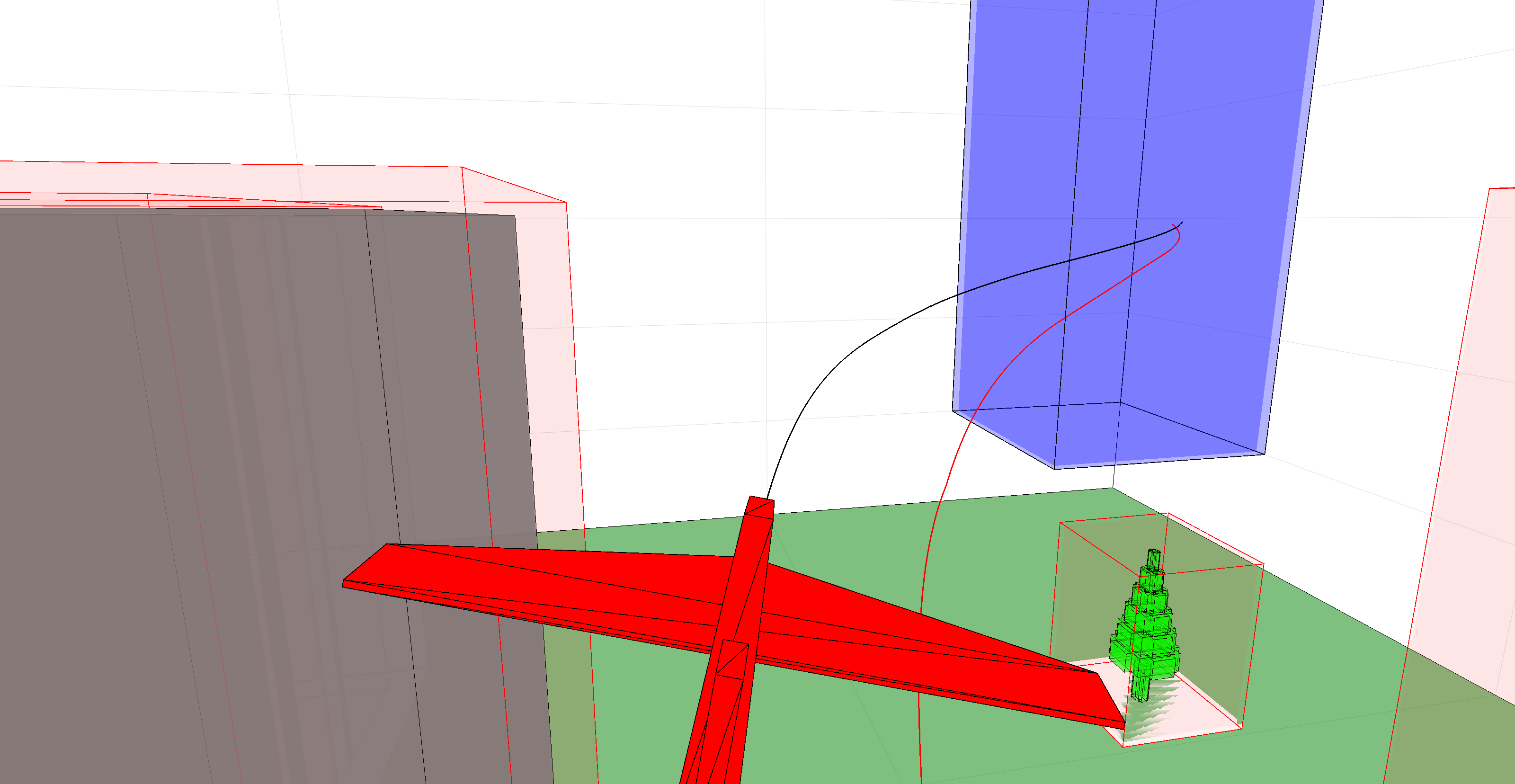}
		\caption{}
		\label{fig:plane}
	\end{subfigure}	
	\caption{{Left:} snapshot of the nominal Dubins motion plan (red), along with the bounding ellipsoidal tube. The actual trajectory (black) stays within the ellipsoidal tube at all times. {Right:} Close-up of a tight turn at the end towards the goal.\vspace{-2em}}
	\label{fig:Plane_setup}
\end{figure}

For an application of this bound to online motion planning, we created a cluttered obstacle environment with trees and buildings (see Figure~\ref{fig:plane_env}). The goal region is the blue shaded box in the corner, and the plane starts at position $(1,1,5)$. The nominal motion plan was computed using kinodynamic FMT* \cite{SchmerlingJansonEtAl2015}, with the locally optimal trajectories in~\cite{ChitsazLaValle2007} as steering connections). Having computed all steering connections offline (< 2 min computation time for 5000 samples), the online computation time for the trajectory was on the order of 100 ms. Collision checking was performed using obstacles inflated by the size of the plane (roughly a $2\times 2\times 0.5$m box envelope) and the projection of the ellipsoidal bound $\teb$ onto $(x_r, \, y_r, \, z_r)$ (rotated and translated by the orientation and position of the Dubins plane). 
In Figure~\ref{fig:plane_env}, the nominal motion plan (i.e., the plan for the decoupled Dubins model) is shown in red, while the actual trajectory followed by the plane is shown in black. 
We overlay the projection of set $\teb$ onto position space, and its sweep along the nominal Dubins airplane motion plan. The snapshot confirms that this ``tube'' remains obstacle free, and highlights the necessity of using a $\teb$-based buffer as the plane negotiates difficult turns (see Figure~\ref{fig:plane}). 
A video of this simulation can be found at \url{https://www.youtube.com/watch?v=UsfaWsjectQ}. \revision{Additional plots and simulations are provided in Appendix~\ref{sec:add}.}


\vspace{-4mm}
\section{Conclusions}\label{sec:conclusion}\vspace{-.5em}
In this work we have presented a principled and scalable approach for planning with model mismatch. Specifically, by harnessing the tool of SOS programming, we designed an algorithmic framework to compute, offline, for a given pair of planning and tracking models, a feedback tracking controller and associated tracking bound; this bound can then used by the planning algorithm as a safety-margin when generating motion plans.  The efficacy of the approach was verified via illustrative examples, which include a comparison with a state-of-the-art method based on reachability analysis. 

There are several key avenues for future work: First, we would like to investigate structural conditions that guarantee fulfillment of Assumption \ref{assum:relative}, along with augmenting the catalog of planning/tracking models in Table 1. Second, we plan to investigate how the conservativeness of our approach may be reduced by using more complex certificates such as the Stengle Positivstellensatz. Third, to fully explore the scalability advantages of our approach, we will consider more tractable relaxations of SOS programming, such as SDSOS and DSOS optimization~\cite{AhmadiMajumdar2014}. Fourth, it is of interest to study how to minimize the \emph{projected} volume of the TEB onto the error states of interest.  
Finally, we plan on implementing our approach on agile robotic systems such as unmanned aerial vehicles.\vspace{-1em}






\renewcommand{\baselinestretch}{0.85}
\bibliographystyle{splncs03}

\begin{thebibliography}{10}
\providecommand{\url}[1]{\texttt{#1}}
\providecommand{\urlprefix}{URL }

\bibitem{AhmadiMajumdar2014}
Ahmadi, A.A., Majumdar, A.: {DSOS} and {SDSOS} optimization: {LP} and
  {SOCP}-based alternatives to sum of squares optimization. In: {IEEE Annual
  Conf.\ on Information Sciences and Systems} (2014)

\bibitem{AhmadiMajumdar2016}
Ahmadi, A.A., Majumdar, A.: Some applications of polynomial optimization in
  operations research and real-time decision making. {Optimization Letters}
  20(4),  709--729 (2016)

\bibitem{ApS2017}
{ApS}, M.: {MOSEK} optimization software (2017), {Available at
  }\url{https://mosek.com/}

\bibitem{Ben-TalNemirovski2001}
Ben-Tal, A., Nemirovski, A.: Lectures on modern convex optimization: analysis,
  algorithms, and engineering applications. {SIAM} (2001)

\bibitem{BurridgeRizziEtAl1999}
Burridge, R.R., Rizzi, A.A., Koditschek, D.E.: Sequential composition of
  dynamically dexterous robot behaviors. {Int.\ Journal of Robotics Research}
  18(6),  534--555 (1999)

\bibitem{ChenBansalEtAl2018}
Chen, M., Bansal, S., Fisac, J.F., Tomlin, C.J.: Robust sequential path
  planning under disturbances and adversarial intruder. {IEEE Transactions on
  Control Systems Technology}  (2018), in press

\bibitem{ChenHerbertEtAl2016b}
Chen, M., Herbert, S., Tomlin, C.J.: Fast reachable set approximations via
  state decoupling disturbances. In: {Proc.\ IEEE Conf.\ on Decision and
  Control} (2016)

\bibitem{ChenHerbertEtAl2018}
Chen, M., Herbert, S.L., Vashishtha, M.S., Bansal, S., Tomlin, C.J.:
  Decomposition of reachable sets and tubes for a class of nonlinear systems.
  {IEEE Transactions on Automatic Control}  (2018), in press

\bibitem{ChitsazLaValle2007}
Chitsaz, H., LaValle, S.M.: Time-optimal paths for a dubins airplane. In:
  {Proc.\ IEEE Conf.\ on Decision and Control} (2007)

\bibitem{CoddingtonLevinson1955}
Coddington, E.A., Levinson, N.: Theory of Ordinary Differential Equations.
  {McGraw-Hill} (1955)

\bibitem{HerbertChenEtAl2017}
Herbert, S.L., Chen, M., Han, S., Bansal, S., Fisac, J.F., Tomlin, C.J.:
  {FaSTrack:} a modular framework for fast and guaranteed safe motion planning.
  In: {Proc.\ IEEE Conf.\ on Decision and Control} (2017)

\bibitem{IwasakiHara2005}
Iwasaki, T., Hara, S.: Generalized {KYP} lemma: unified frequency domain
  inequalities with design applications. {IEEE Transactions on Automatic
  Control}  50(1),  41--59 (2005)

\bibitem{JansonSchmerlingEtAl2015}
Janson, L., Schmerling, E., Clark, A., Pavone, M.: {Fast} {Marching} {Tree:} a
  fast marching sampling-based method for optimal motion planning in many
  dimensions. {Int.\ Journal of Robotics Research}  34(7),  883--921 (2015)

\bibitem{KaramanFrazzoli2011}
Karaman, S., Frazzoli, E.: Sampling-based algorithms for optimal motion
  planning. {Int.\ Journal of Robotics Research}  30(7),  846--894 (2011)

\bibitem{KousikVaskovEtAl2017}
Kousik, S., Vaskov, S., Johnson-Roberson, M., Vasudevan, R.: Safe trajectory
  synthesis for autonomous driving in unforeseen environments. In: {Proc. ASME
  Dynamic Systems and Control Conf.} (2017)

\bibitem{LaValle2011}
LaValle, S.M.: Motion planning: Wild frontiers. {IEEE Robotics and Automation
  Magazine}  18(2),  108--118 (2011)

\bibitem{LaValleKuffner2001}
LaValle, S.M., Kuffner, J.J.: Randomized kinodynamic planning. {Int.\ Journal
  of Robotics Research}  20(5),  378--400 (2001)

\bibitem{LiLittlefieldEtAl2016}
Li, Y., Littlefield, Z., Bekris, K.E.: Asymptotically optimal sampling-based
  kinodynamic planning. {Int.\ Journal of Robotics Research}  35(5),  528–564
  (2016)

\bibitem{MajumdarAhmadiEtAl2013}
Majumdar, A., Ahmadi, A.A., Tedrake, R.: Control design along trajectories with
  sums of squares programming. In: {Proc.\ IEEE Conf.\ on Robotics and
  Automation} (2013)

\bibitem{MajumdarTedrake2013}
Majumdar, A., Tedrake, R.: Robust online motion planning with regions of finite
  time invariance. In: Algorithmic Foundations of Robotics X. {Springer} (2013)

\bibitem{MajumdarTedrake2017}
Majumdar, A., Tedrake, R.: Funnel libraries for real-time robust feedback
  motion planning. {Int.\ Journal of Robotics Research}  36(8),  947--982
  (2017)

\bibitem{OwenBeardEtAl2015}
Owen, M., Beard, R.W., McLain, T.W.: Implementing dubins airplane paths on
  fixed-wing {UAVs}. In: Handbook of Unmanned Aerial Vehicles. {Springer
  Dordrecht} (2015)

\bibitem{Parrilo2000}
Parrilo, P.A.: Structured Semidefinite Programs and Semialgebraic Geometry
  Methods in Robustness and Optimization. Ph.D. thesis, {Massachusetts Inst.\
  of Technology} (2000)

\bibitem{PosaKoolenEtAl2017}
Posa, M., Koolen, T., Tedrake, R.: Balancing and step recovery capturability
  via sums-of-squares optimization. In: {Robotics: Science and Systems} (2017)

\bibitem{Putinar1993}
Putinar, M.: Positive polynomials on compact semi-algebraic sets. {Indiana
  Univ.\ Mathematics Journal}  42(3),  969--984 (1993)

\bibitem{Rajamani2012}
Rajamani, R.: Vehicle Dynamics and Control. {Springer}, Boston, MA, 2 edn.
  (2012)

\bibitem{RatliffZuckerEtAl2009}
Ratliff, N., Zucker, M., Bagnell, J.A., Srinivasa, S.: {CHOMP:} gradient
  optimization techniques for efficient motion planning. In: {Proc.\ IEEE
  Conf.\ on Robotics and Automation} (2009)

\bibitem{SchmerlingJansonEtAl2015b}
Schmerling, E., Janson, L., Pavone, M.: Optimal sampling-based motion planning
  under differential constraints: the drift case with linear affine dynamics.
  In: {Proc.\ IEEE Conf.\ on Decision and Control} (2015)

\bibitem{SchmerlingJansonEtAl2015}
Schmerling, E., Janson, L., Pavone, M.: Optimal sampling-based motion planning
  under differential constraints: the driftless case. In: {Proc.\ IEEE Conf.\
  on Robotics and Automation} (2015), {Extended version available at
  }\url{http://arxiv.org/abs/1403.2483/}

\bibitem{SchmerlingPavone2017}
Schmerling, E., Pavone, M.: Evaluating trajectory collision probability through
  adaptive importance sampling for safe motion planning. In: {Robotics: Science
  and Systems} (2017)

\bibitem{SchulmanHoEtAl2013}
Schulman, J., Ho, J., Lee, A., Awwal, I., Bradlow, H., Abbeel, P.: Finding
  locally optimal, collision-free trajectories with sequential convex
  optimization. In: {Robotics: Science and Systems} (2013)

\bibitem{SinghMajumdarEtAl2017}
Singh, S., Majumdar, A., Slotine, J.J.E., Pavone, M.: Robust online motion
  planning via contraction theory and convex optimization. In: {Proc.\ IEEE
  Conf.\ on Robotics and Automation} (2017), {Extended Version, Available at
  }\url{http://asl.stanford.edu/wp-content/papercite-data/pdf/Singh.Majumdar.Slotine.Pavone.ICRA17.pdf}

\bibitem{SteinhardtTedrake2012}
Steinhardt, J., Tedrake, R.: Finite-time regional verification of stochastic
  non-linear systems. {Int.\ Journal of Robotics Research}  31(7),  901--923
  (2012)

\bibitem{TanabeChen2018}
Tanabe, K., Chen, M.: {beacls Library}, {Available at
  }\url{https://github.com/HJReachability/beacls}

\bibitem{TedrakeManchesterEtAl2010}
Tedrake, R., Manchester, I.R., Tobenkin, M., Roberts, J.W.: {LQR}-trees:
  Feedback motion planning via sums-of-squares verification. {Int.\ Journal of
  Robotics Research}  29(8),  1038--1052 (2010)

\bibitem{TobenkinPermenterEtAl2013}
Tobenkin, M.M., Permenter, F., Megretski, A.: {SPOTLESS} polynomial and conic
  optimization (2013), software available from
  \url{https://github.com/spottoolbox/spotless}

\end{thebibliography}
\newcommand{\noopsort}[1]{} \newcommand{\printfirst}[2]{#1}
  \newcommand{\singleletter}[1]{#1} \newcommand{\switchargs}[2]{#2#1}

\newpage
\renewcommand{\baselinestretch}{0.91}
\appendix
\begin{changemargin}{-1cm}{-1cm}
\section*{Appendix}
\section{Catalog of Planning/Tracking Models}\label{sec:table}
\begin{table}[h!] \scriptsize
\begin{center}
\scalebox{0.77}{
\Rotatebox{90}{%
\begin{tabular}{ | l | c | l | c | c | c | c | } \hline
Tracking system 
    & Tracking model 
    & Planning system
    & Planning model 
    & Transformation $\rtrans$
    & Relative system dynamics 
    & \parbox[t]{2cm}{Suggested bounded state mapping $c(r)$} \\ \hline
\parbox{2.5cm}{\textbf{4D car} \\$(x,y)$ -- position \\ $\theta$ -- heading \\ $v$ -- speed \\ $\ctrl_\omega$ -- turn rate control \\ $\ctrl_a$ -- accel. control}
    & $\begin{bmatrix} \dot x \\ \dot y \\ \dot \theta \\ \dot v \end{bmatrix} 
        = \begin{bmatrix} v \cos \theta \\ v \sin \theta \\ u_\omega \\ u_a \end{bmatrix}$ 
    & \parbox{2cm}{\textbf{2D single integrator} \\ $(\hat x,\hat y)$ -- position \\ $(\hat \ctrl_{\hat x}, \hat \ctrl_{\hat x})$ \\ -- speed control}
    & $\begin{bmatrix} \dot{\hat x} \\ \dot{\hat y} \end{bmatrix} 
        = \begin{bmatrix} \hat u_{\hat x} \\ \hat u_{\hat y} \end{bmatrix}$ 
    & $\begin{bmatrix} R(\theta) & \zero_{2\times2} \\ \zero_{1\times2} & \begin{bmatrix} 0 & 1\end{bmatrix}\end{bmatrix}$ 
    & $\begin{aligned} 
        \begin{bmatrix} \dot x_r \\ \dot y_r \\ \dot v \end{bmatrix}
            = \begin{bmatrix} v + u_{\omega} y_r - \tilde u_{\hat x} \\ -u_{\omega} x_r - \tilde u_{\hat y} \\ u_{a} \end{bmatrix},  \begin{bmatrix} \tilde u_{\hat x} \\ \tilde u_{\hat y} \end{bmatrix} 
            =R(\theta)
            \begin{bmatrix} \hat u_{\hat x} \\\hat u_{\hat y} \end{bmatrix}
        \end{aligned}$ 
    & $\eye_3$ \\ \hline
\parbox{2.5cm}{\textbf{5D car} \\ $(x,y)$ -- position \\ $\theta$ -- heading \\ $v$ -- speed \\ $\omega$ -- turn rate \\ $\ctrl_a$ -- accel. control \\ $\ctrl_\alpha$ -- ang. accel. control} 
    & $\begin{bmatrix} \dot x\\ \dot y\\ \dot\theta\\ \dot v\\ \dot \omega \end{bmatrix} 
        = \begin{bmatrix} v \cos \theta \\ v \sin \theta\\ \omega \\ \ctrl_a\\ \ctrl_\alpha \end{bmatrix}$ 
    & \parbox{2cm}{\textbf{3D Dubins car} \\ $(\hat x,\hat y)$ -- position \\ $\hat\theta$ -- heading \\ $\hat v$ -- constant speed \\ $\hat\ctrl_\omega$ \\ -- turn rate control}
    &$\begin{bmatrix} \dot {\hat x}\\ \dot {\hat y}\\ \dot {\hat \theta} \end{bmatrix}
        = \begin{bmatrix} \hat v \cos \hat\theta \\ \hat v \sin \hat\theta\\ \hat \ctrl _ {\hat\omega}\end{bmatrix}$ 
    & $\begin{bmatrix} R(\hat\theta) & \mathbf{0_{2\times 3}} \\ \mathbf{0_{3\times 2}} & \mathbf I_3\end{bmatrix}$ 
    & $ \begin{bmatrix} \dot x_r\\ \dot y_r\\ \dot\theta_r\\ \dot v\\  \dot \omega \end{bmatrix}
        = \begin{bmatrix} - \hat v + v \cos \theta_r + \hat\ctrl_ {\hat\omega} y_r\\ v \sin \theta_r - \hat\ctrl_ {\hat\omega} x_r\\ \omega - \hat \ctrl_{\hat\omega} \\ \ctrl_a\\\ctrl_\alpha \end{bmatrix}$ 
    & $\begin{bmatrix} x_r \\ y_r \\ \theta_r \\ v \cos \theta_r - \hat v \\ v \sin \theta_r \\ \omega \end{bmatrix}$ \\ \hline
\parbox{2.5cm}{\textbf{6D planar quadrotor} \\ $(x,z)$ -- position \\ $(v_x,v_z)$ -- velocity \\ $\theta$ -- pitch \\ $\omega$ -- pitch rate \\ $\ctrl_T$ -- thrust control \\ $\ctrl_\tau$ \\ -- ang. accel. control}
    & $\begin{bmatrix} \dot x \\ \dot z \\ \dot v_x \\ \dot v_z \\ \dot \theta \\ \dot \omega \end{bmatrix} 
        = \begin{bmatrix} v_x \\ v_z \\ -u_{T}\sin\theta \\ u_{T}\cos\theta - g \\ \omega \\ u_{\tau} \end{bmatrix}$ 
    & \parbox{2cm}{\textbf{4D double integrator} \\ $(\hat x,\hat z)$ -- position \\ $(\hat v_x,\hat v_z)$ -- velocity \\ $(\hat\ctrl_{\hat x}, \hat\ctrl_{\hat y})$ \\ -- accel. control}
    & $\begin{bmatrix} \dot {\hat x} \\ \dot{\hat z} \\ \dot {\hat v}_x \\ \dot{\hat v}_z \end{bmatrix} 
        = \begin{bmatrix} \hat v_x \\ \hat v_z \\ \hat u_{\hat x} \\ \hat u_{\hat z} \end{bmatrix}$ 
    & $\eye_6$ 
    & $\begin{bmatrix} \dot x _r\\ \dot{z}_r \\ \dot v_{x,r} \\  \dot v_{z,r} \\ \dot \theta \\ \dot \omega \end{bmatrix} 
        = \begin{bmatrix} v_{x,r} \\ v_{z,r} \\ -u_{T}\sin\theta - \hat{u}_{\hat x} \\ u_{T}\cos\theta - g - \hat{u}_{\hat y} \\ \omega \\ u_{\tau} \end{bmatrix}$ 
    & $\eye_6$ \\ \hline        
\parbox{2.5cm}{\textbf{8D plane}, see \cite{SchmerlingPavone2017} \\ $(x,y,z)$ -- position \\ $\psi$ -- heading\\ $v$ -- speed \\ $\gamma$ -- flight path angle \\ $\phi$ -- roll \\ $\alpha$ -- angle of attack \\ $(\ctrl_{a}, \ctrl_{\dot\phi}, \ctrl_{\dot\alpha})$ -- control}
    & $\begin{bmatrix} \dot x \\ \dot y \\ \dot z \\ \dot \psi \\ \dot v \\ \dot \gamma \\ \dot\phi \\ \dot\alpha \end{bmatrix} 
        = \begin{bmatrix} v \cos \psi \cos \gamma \\ v \sin \psi \cos \gamma \\ v \sin \gamma \\ -\frac{F_\text{lift}(v,\alpha)\sin\phi}{mv\cos\gamma} \\ u_a - \frac{F_\text{drag}(v, \alpha)}{m} - g \sin \gamma \\ \frac{F_\text{lift}(v,\alpha)\cos\phi}{mv}  - \frac{g\cos\gamma}{v} \\ u_{\dot\phi} \\ u_{\dot\alpha} \end{bmatrix}$ 
    & \parbox{2cm}{\textbf{4D plane} \\ $(\hat x, \hat y, \hat z)$ -- position \\ $\hat \psi$ -- heading \\ $\hat v$ -- constant speed \\ $\hat \ctrl_{\hat \omega}$ -- turn rate control \\ $\ctrl_{\hat v_z}$ \\ -- vert. speed control}
    & $\begin{bmatrix} \dot {\hat x} \\ \dot {\hat y} \\ \dot{\hat z} \\ \dot{\hat \psi} \end{bmatrix} 
        = \begin{bmatrix} \hat v \cos \hat\psi \\ \hat v \sin \hat\psi \\ \ctrl_{\hat\omega} \\ \ctrl_{\hat v_z} \end{bmatrix}$ 
    & $\begin{bmatrix} R(\hat\psi) & \mathbf{0_{2\times 6}} \\ \mathbf{0_{6\times 2}} & \mathbf{I_{6}} \end{bmatrix}$ 
    & $\begin{bmatrix} \dot x_r\\ \dot y_r\\ \dot z_r \\ \dot \psi_r \\ \dot v \\ \dot \gamma \\ \dot \phi \\ \dot \alpha \end{bmatrix} 
        = \begin{bmatrix} v \cos \psi_r \cos \gamma - \hat v + \hat\omega y_r \\ v \sin\psi_r \cos\gamma - \hat\omega x_r \\ v\sin\gamma - \hat v_z \\ -\frac{F_\text{lift}(v,\alpha)\sin\phi}{mv\cos\gamma} - \hat\omega\\ u_a - \frac{F_\text{drag}(v, \alpha)}{m} - g \sin \gamma \\ \frac{F_\text{lift}(v,\alpha)\cos\phi}{mv}  - \frac{g\cos\gamma}{v} \\ u_{\dot\phi} \\ u_{\dot\alpha} \end{bmatrix}$
    &  $\begin{bmatrix} 
    x_r \\ y_r \\  z_r \\ \psi_r \\ v-\hat{v} \\  \gamma \\  \phi \\ \alpha \end{bmatrix}$ \\ \hline
\parbox{2.5cm}{\textbf{6D bicycle}, see \cite[p. 27]{Rajamani2012} \\ $(X,Y)$ -- position \\ $\psi$ -- heading \\ $(v_x, v_y)$ \\ -- body frame velocity \\ $\omega$ -- turn rate}
    &$\begin{bmatrix} \dot X \\ \dot Y \\ \dot \psi \\ \dot v_x \\ \dot v_y \\ \dot \omega \\ \end{bmatrix} 
        = \begin{bmatrix} v_x \cos \psi - v_y \sin \psi \\ v_x \sin \psi + v_y \cos \psi \\ \omega \\ \omega v_y + a_x \\ -\omega v_x + \frac{2}{m}(F_{c,f} \cos \delta_f + F_{c,r}) \\ \frac{2}{I_z}(l_f F_{c,f} - l_r F_{c,r}) \end{bmatrix}$
    & \parbox{2cm}{\textbf{3D Dubins car} \\ $(\hat X, \hat Y)$ -- position \\ $\hat \psi$ -- heading \\ $\hat v$ -- constant speed \\ $\hat \omega$ -- turn rate control}
    & $\begin{bmatrix} \dot {\hat X} \\ \dot {\hat Y} \\ \dot{\hat \psi} \end{bmatrix} 
        = \begin{bmatrix} \hat v \cos \hat\psi \\ \hat v \sin \hat\psi \\ \hat\omega \end{bmatrix}$ 
    & $\begin{bmatrix} R(\hat\psi) & \mathbf{0_{2\times 4}} \\ \mathbf{0_{4\times 2}} & \mathbf{I_{4}} \end{bmatrix}$ 
    & $\begin{bmatrix} \dot X_r \\ \dot Y_r \\ \dot \psi_r \\ \dot v_x \\ \dot v_y \\ \dot \omega \end{bmatrix} 
        = \begin{bmatrix} v_x \cos \psi_r - \hat v + \hat \omega Y_r \\ v_x \sin \psi_r - \hat \omega X_r + v_y \\ \omega - \hat \omega \\ \omega v_y + a_x \\ -\omega v_x + \frac{2}{m}(F_{c,f} \cos \delta_f + F_{c,r}) \\ \frac{2}{I_z}(l_f F_{c,f} - l_r F_{c,r}) \end{bmatrix}$
    & $\eye_6$ \\ \hline
\parbox{2.5cm}{\textbf{10D near-hover quadrotor}, see \cite{ChenHerbertEtAl2018} \\ $(x,y,z)$ -- position \\ $(v_x, v_y, v_z)$ -- velocity\\ $(\theta_x, \theta_y)$ -- pitch and roll\\ $(\omega_x, \omega_y)$ \\ -- pitch and roll rates \\ $u_z$ -- thrust control \\ $(u_x, u_y)$ \\ -- pitch and roll control}
    & $\begin{bmatrix} \dot{x}\\ \dot{y}\\ \dot{z}\\ \dot{v_x} \\ \dot{v_y}\\ \dot{v_z}\\ \dot{\theta_x}\\ \dot{\theta_y}\\ \dot\omega_x\\ \dot\omega_y \end{bmatrix}
        = \begin{bmatrix} v_x \\ v_y\\ v_z\\ g \tan \theta_x\\ g \tan \theta_y\\ k_T u_z - g \\ -d_1 \theta_x + \omega_x\\ -d_1 \theta_y + \omega_y\\ -d_0 \theta_x + n_0 u_x \\ -d_0 \theta_y + n_0 u_y \end{bmatrix}$ 
    & \parbox{2cm}{\textbf{3D single integrator} \\ $(\hat x, \hat y, \hat z)$ -- position \\ $((\hat\ctrl_{\hat x}, \hat\ctrl_{\hat y}, \hat \ctrl_{\hat z})$ \\ -- speed control}        
    & $\begin{bmatrix} \dot{\hat x}\\ \dot{\hat y}\\ \dot{\hat z}\\ \end{bmatrix} 
        = \begin{bmatrix} \hat\ctrl_{\hat x} \\ \hat\ctrl_{\hat y} \\ \hat \ctrl_{\hat z} \end{bmatrix}$ 
    & $\eye_{10}$ 
    & $\begin{bmatrix} \dot x_r\\ \dot v_x\\ \dot \theta_x\\ \dot\omega_x\\ \dot y_r\\ \dot v_y\\ \dot \theta_y\\ \dot\omega_y\\ \dot z_r\\ \dot v_z \end{bmatrix} 
        = \begin{bmatrix} v_x - \hat\ctrl_{\hat x} \\ g \tan \theta_x\\ d_1 \theta_x + \omega_x\\ -d_0 \theta_x + n_0 u_x\\ v_y - \hat\ctrl_{\hat y}\\ g \tan \theta_y\\ -d_1 \theta_y + \omega_y\\ -d_0 \theta_y + n_0 u_y\\ v_z - \hat\ctrl_{\hat z}\\ k_T u_z - g \end{bmatrix}$ 
    & $\eye_{10}$ \\ \hline
    \end{tabular}
    }
    }
  \end{center}
\caption{{Catalog of tracking and planning system models; $R(\cdot) = \protect\begin{bmatrix}\cos(\cdot) & \sin(\cdot) \\ -\sin(\cdot) & \cos(\cdot)\protect\end{bmatrix}$ denotes the rotation matrix.} \label{tab:cat}}  
\end{table}
\end{changemargin}

\section{Additional Numerical Experiments} \label{sec:add}
\revision{
\subsection{Comparison with the HJ Method}

As an additional comparison to the HJ-based approach in \cite{HerbertChenEtAl2017}, we consider the case where a dynamically extended 4D Dubins car model is used to track a 2D single-integrator model (that is, the first example in Table \ref{tab:cat}). 
The error and auxiliary states are $e = (x_r, y_r)$, and $\eta = v$, respectively. Note that the map $\phi(\cdot)$ for this system, as given in Table \ref{tab:cat}, exploits the rotational invariance of the tracking model, thereby yielding a dimensionality reduction. In addition, given a 2-norm bound on $\|(\hat{u}_{\hat{x}}, \hat{u}_{\hat{y}})\|_2$ and the norm-preservation property of rotations, the same bound holds for $(\tilde{u}_{\hat x}, \tilde{u}_{\hat y})$. Finally, since the rotation angle $\theta$ is collapsed in the relative system state definition through the rotation matrix $R(\theta)$, following the computation of the bound $\teb$ for the relative system state $(x_r,y_r,v)$, we can project this bound back to the space $(e',v)$ where $e' = (x-\hat{x},y-\hat{y})$ is the absolute error in relative position, by simply taking the union
$\small
\teb' := \bigcup_{v} \bigcup_{\theta \in [-\pi,\pi]} \left\{ 
 \begin{bmatrix} R(\theta)^T e  \\ v \end{bmatrix}  : \begin{bmatrix} e  \\ v\end{bmatrix} \in \teb \right\}.
$
For a fixed $v$, this is a rotational sweep of the projection of the set $\teb$ onto $e$. Since the true error bound we care about is $e'$, all results below will be displayed with respect to $\teb'$.

For this example, we used the following control bounds: $(u_a, u_{\omega}) \in [-1,1] \text{ m/s}^2 \times [-1,1] \text{ rad/s}$, and $\| (\hat{u}_{\hat x}, \hat{u}_{\hat y}) \| \leq 0.1$m/s. 
  
The bounded state mapping $c(\cdot)$ was the identity. We parameterized the feedback controller and function $\vft$ up to degree $4$ and initialized the infeasible start algorithm (i.e., Algorithm~\ref{alg:infeas}) with $\vft(\rt) = [\rt, \rt^2]  Q [\rt, \rt^2]^T$ where $\rt^2$ denotes component-wise square of $\rt$, and $Q$ was a randomly generated psd matrix. Over 10 initializations, on average, 7 iterations were needed to exit Algorithm~\ref{alg:infeas}. The optimization loop in Algorithm~\ref{alg:feas} took <90s on all runs for convergence.
\begin{figure}[h]
\centering
\includegraphics[width=\textwidth,clip]{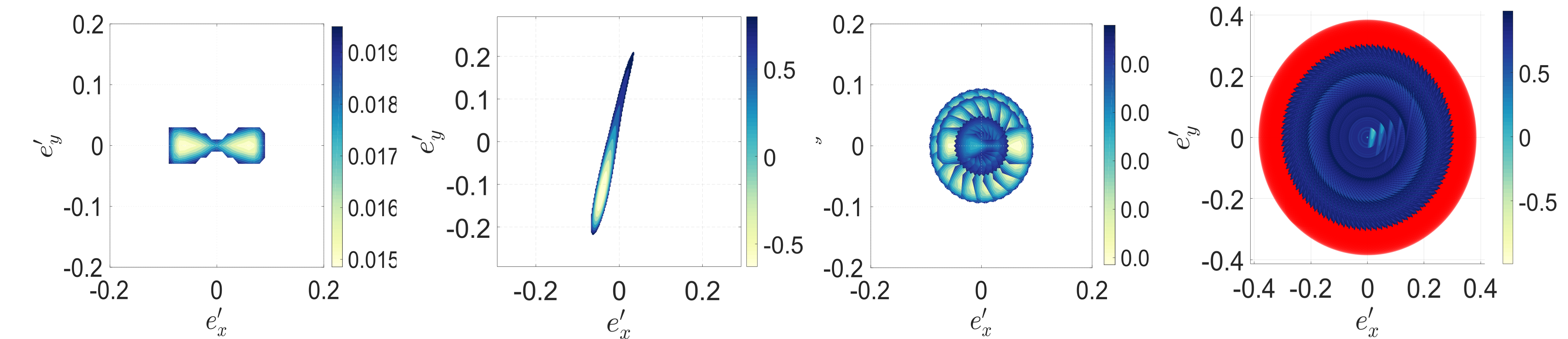}
\caption{Left two plots: Projection of the invariant set $\teb'$ onto $e'$ for the fixed slice $(v,\theta) = (0.013,0)$ (left: HJ, right: SOS); Right two plots: Projection of the invariant set $\teb'$ onto $e'$ for all $(v,\theta)$ (left: HJ, right: SOS). In red is the projection of the bounding ellipsoid $\elps$ onto $e'$. }
	\label{fig:dubins_V}
\end{figure}

Figure~\ref{fig:dubins_V} illustrates the results of the HJ and SOS solutions. Specifically, in the left two plots of Figure~\ref{fig:dubins_V}, we plot the projection of the set $\teb'$ onto $e'$ for a fixed value of $(v,\theta) = (\bar{v},\bar{\theta})$, i.e. $\small
	\left\{ R(\bar{\theta})^T e :  \begin{bmatrix} e  \\ \bar{v}\end{bmatrix} \in \teb \right \}$. In the right two plots, we plot the projection of $\teb'$ onto $e'$ over all $(v,\theta)$, i.e. $\small
	\bigcup_{v} \bigcup_{\theta \in [-\pi,\pi]}\left\{ R(\theta)^T e :  \begin{bmatrix} e  \\ v\end{bmatrix} \in \teb \right \}.$ For the SOS solution, the range of $v$ for this union was extracted from the projection of the bounding ellipsoid $\elps$ onto $v$. The projection of this ellipsoid onto $e'$ is shown in red in the background of the rightmost panel of Figure~\ref{fig:dubins_V}, and can be seen to envelop $\teb'$ as expected.  

We note that the SOS solution is within a reasonable ballpark of the optimal solution from HJ. In particular, the projection of $\teb'$ onto $e'$ is approximately a 0.1m radius $l_2$ ball, while the corresponding solution from SOS is a 0.3m radius $l_2$ ball. Similar to the first example in the main text, in order to ensure a fair comparison, the cost function for the min-max game for the HJ method was $e_x + e_y^2 + v^2$, which is the closest analogue to the SOS objective of minimizing the entire volume of $\teb$. The non-smooth nature of the optimal solution as illustrated in the left-most figure in Figure~\ref{fig:dubins_V} illustrates the primary limitation of the SOS method in its ability to approximate such challenging functions.
}
\subsection{Additional High-Dimensional Example}

As an additional high-dimensional example, we consider the case where a  6D planar quadrotor model is used to track a 4D double integrator model  (that is, the third example in Table \ref{tab:cat}). The bounds on the planar quadrotor's inputs are generated by linearly transforming the bounds for each thruster, chosen to be in the  range $[0.1, 1.0] g$ (where $g$ is the gravitational acceleration), by using the model parameters in~\cite{SteinhardtTedrake2012}, namely thruster moment arm 0.25 m, mass 0.486 kg, and rotational inertia 0.00383 $\text{kg m}^2$. The control input for the double integrator is the acceleration $(\hat u_{\hat x}, \hat u_{\hat z})$, with bounds chosen as $\|(\hat u _{\hat x}, \hat u_{\hat z} ) \| \leq 0.5$ m/s$^2$. 
 
The bounded state mapping $c(\cdot)$ is the identity. We parameterize the $K$ and $V$ functions as polynomials up to degree $2$. The trigonometric terms are again approximated using Chebyshev expansions up to degree 3 over the range $\theta \in [-\pi/3, \,\pi/3]$ rad and enforced using the additional constraint as in Section~\ref{subsec:HJ}. The algorithm is initialized using the solution to the Riccati equation for the linearized system at $r=0$; the overall computation time was 172s. The projection onto the $(x_r,z_r)$ dimensions is an ellipse with span $[-0.36, \,0.36] \times [-0.03, \,0.03]$ m, which is quite tight. The projections onto the other dimensions are: $(\dot{x}_r,\dot{z}_r): [-0.42, \,0.42] \times [-0.03, \,0.03]$ m/s, and $(\theta,\omega): [-5.16, \,5.16]^o \times [-18.11, \,18.11]^o$/s.  
\begin{figure}
\centering
	\begin{subfigure}[t]{0.33\textwidth}
		\centering
		\includegraphics[width=\textwidth,clip]{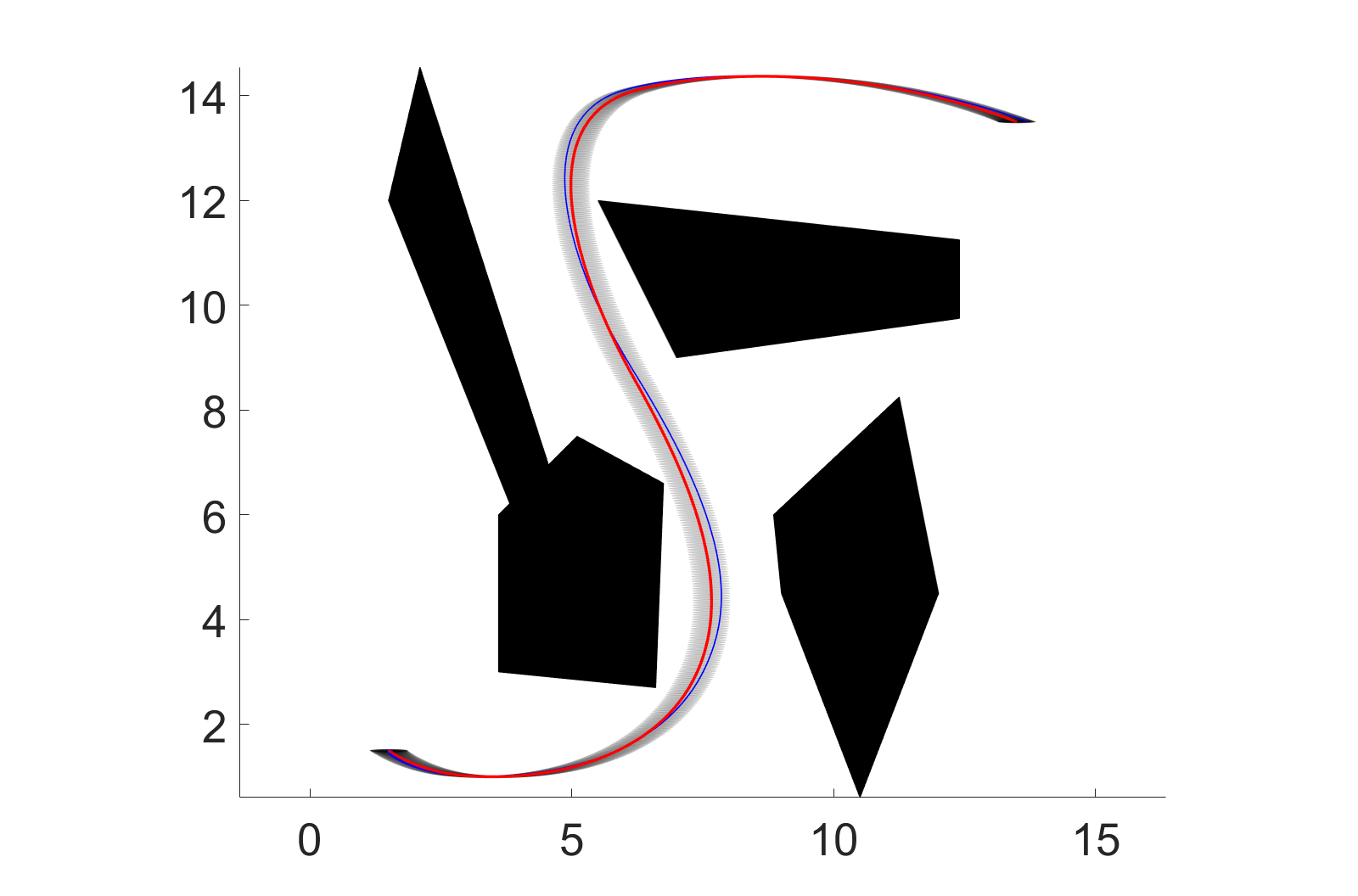}
		\caption{}
		\label{fig:PVTOL_path}
	\end{subfigure}\hspace{-0.5em}
	\begin{subfigure}[t]{0.33\textwidth}
		\centering
		\includegraphics[width=\textwidth,clip]{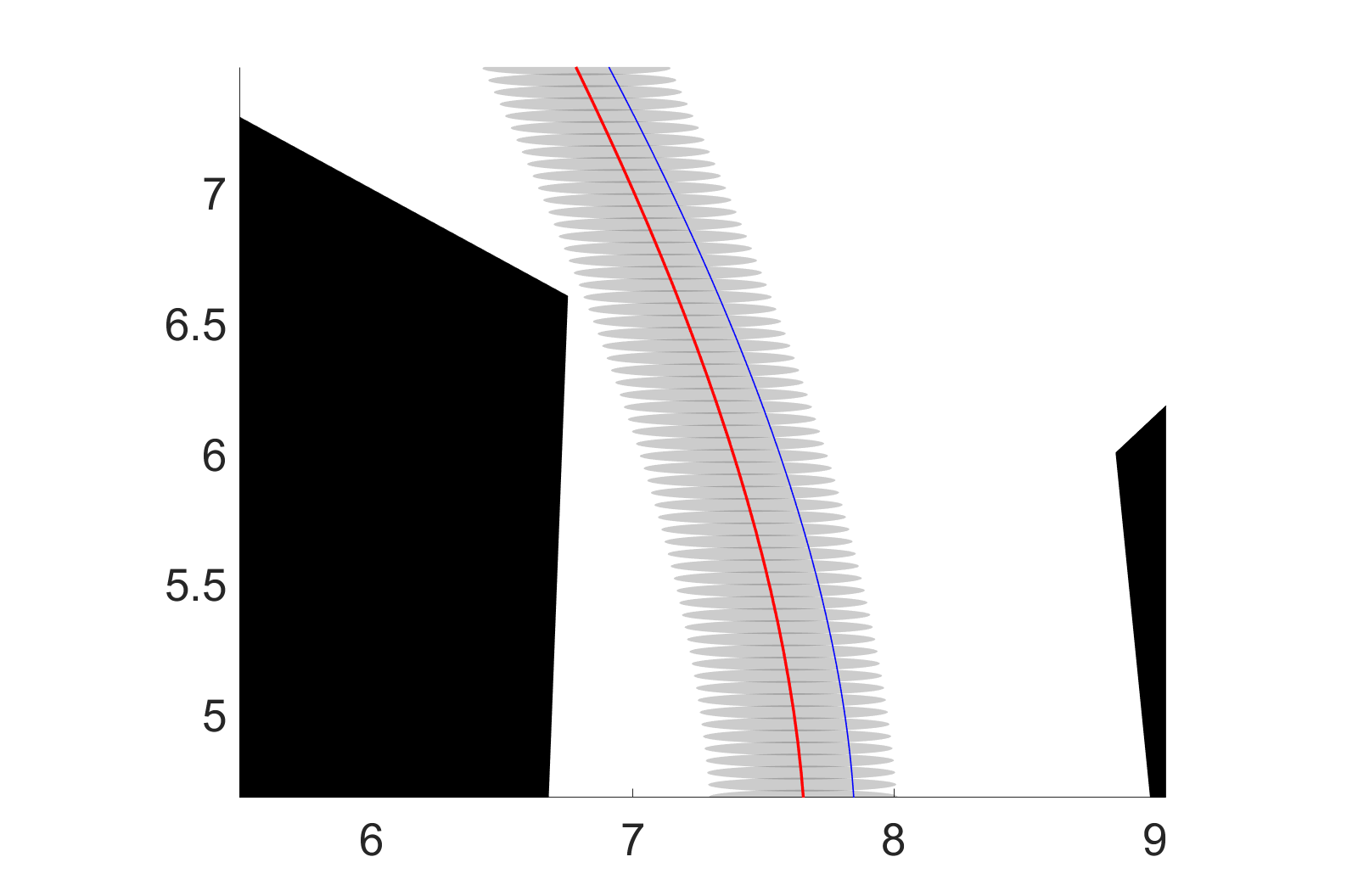}
		\caption{}
		\label{fig:PVTOL_zoom}
	\end{subfigure}	
	\begin{subfigure}[t]{0.33\textwidth}
		\centering
		\includegraphics[width=\textwidth,clip]{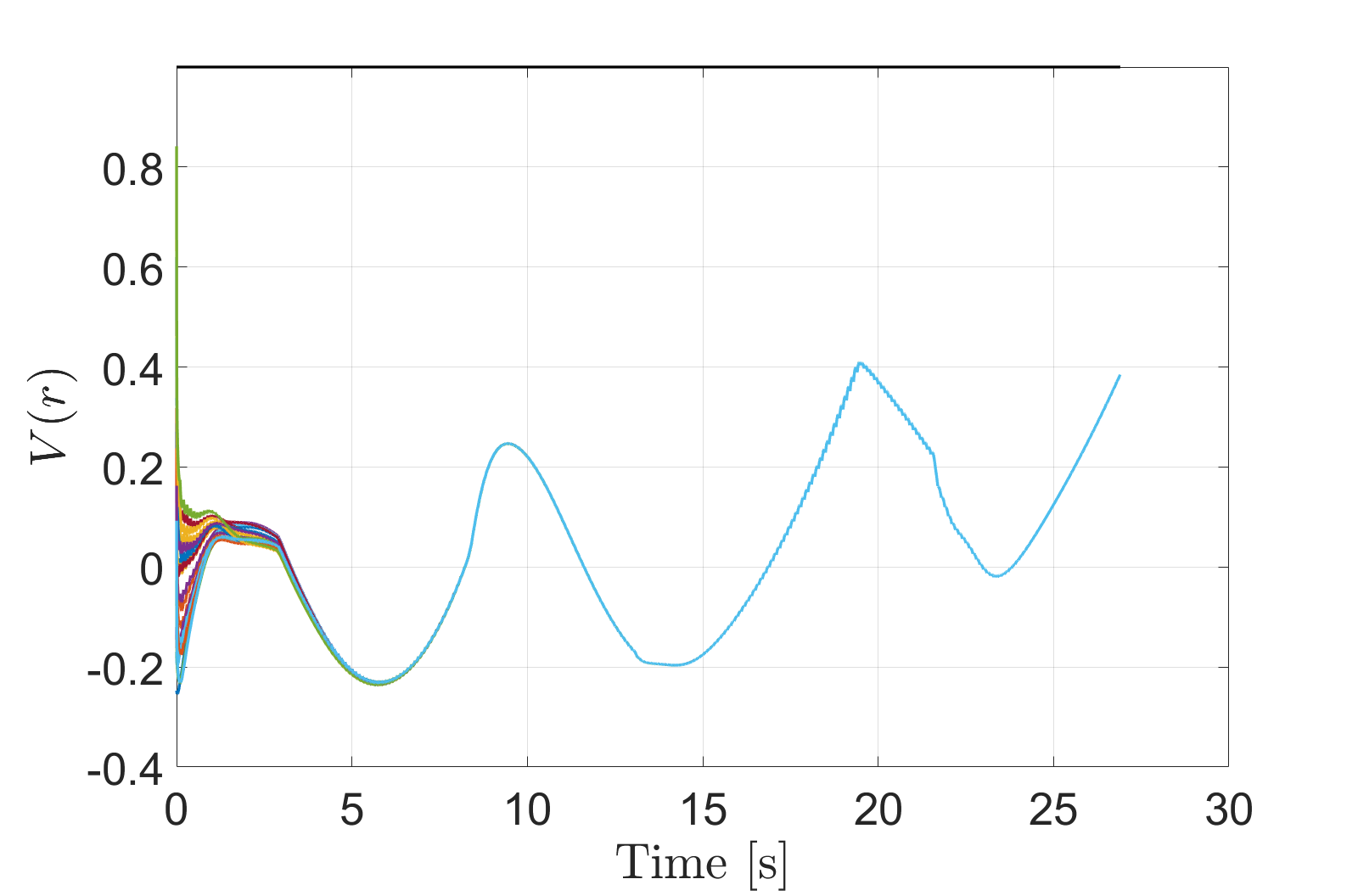}
		\caption{}
		\label{fig:PVTOL_V_t}
	\end{subfigure}
	\caption{(a) Nominal double-integrator motion plan (red), and projection of the set $\teb$ onto the $(x_r, \, z_r)$ dimensions as a safety margin; (b) Zoomed-in view of the tightness of the realized trajectories (blue) with respect to the nominal plan (red) while navigating closely next to an obstacle; (c) Verification that $V(r(t))$ stays below $1$ as required.}
	\label{fig:PVTOL_V_SOS}
\end{figure}

Using the projection of the set $\teb$ onto $(x_r, \,z_r)$ as a safety margin, we planned a nominal motion plan for the obstacle environment depicted in Figure~\ref{fig:PVTOL_path}, by using kinodynamic (double-integrator) FMT*~\cite{SchmerlingJansonEtAl2015b} (red curve). Note that planning using the simple double-integrator model is orders of magnitude (indeed, efficiently implementable in real-time) faster than using non-linear programming methods to find feasible trajectories for the actual planar quadrotor system. 
Superimposed on the plot are the realized trajectories (in blue). The feedback tracking controller ensures that the realized trajectory lies within the swept ellipsoidal tube centered at the nominal motion plan for all time. Figure~\ref{fig:PVTOL_V_t} plots the time series $\vf(t) = \vf(\rstate(t))$; as expected, $\vf(\rstate(t)) \leq 1$ for all $t\geq 0$. 


%

\subsection{Additional Simulations for the 8D Plane}

\revision{
We provide some additional plots and simulations for the 8D Plane tracking a 4D decoupled Dubins planner introduced in Section~\ref{subsec:num_hi}. Figure~\ref{fig:plane_err_pos} plots the error states time series $e = (x_r, y_r, z_r, \psi_r)$ for the nominal planned path in Figure~\ref{fig:plane_env} for a set of varying initial conditions. As expected, the majority of the error is in the $x_r$ direction where the bound is weakest. Figure~\ref{fig:plane_err_V} plots the time series $\vf(\rstate(t))$. As expected, $\vf(\rstate(t)) \leq 1$ for all $t\geq 0$. 
\begin{figure}
\centering
	\begin{subfigure}[t]{0.48\textwidth}
		\centering
		\includegraphics[width=\textwidth,clip]{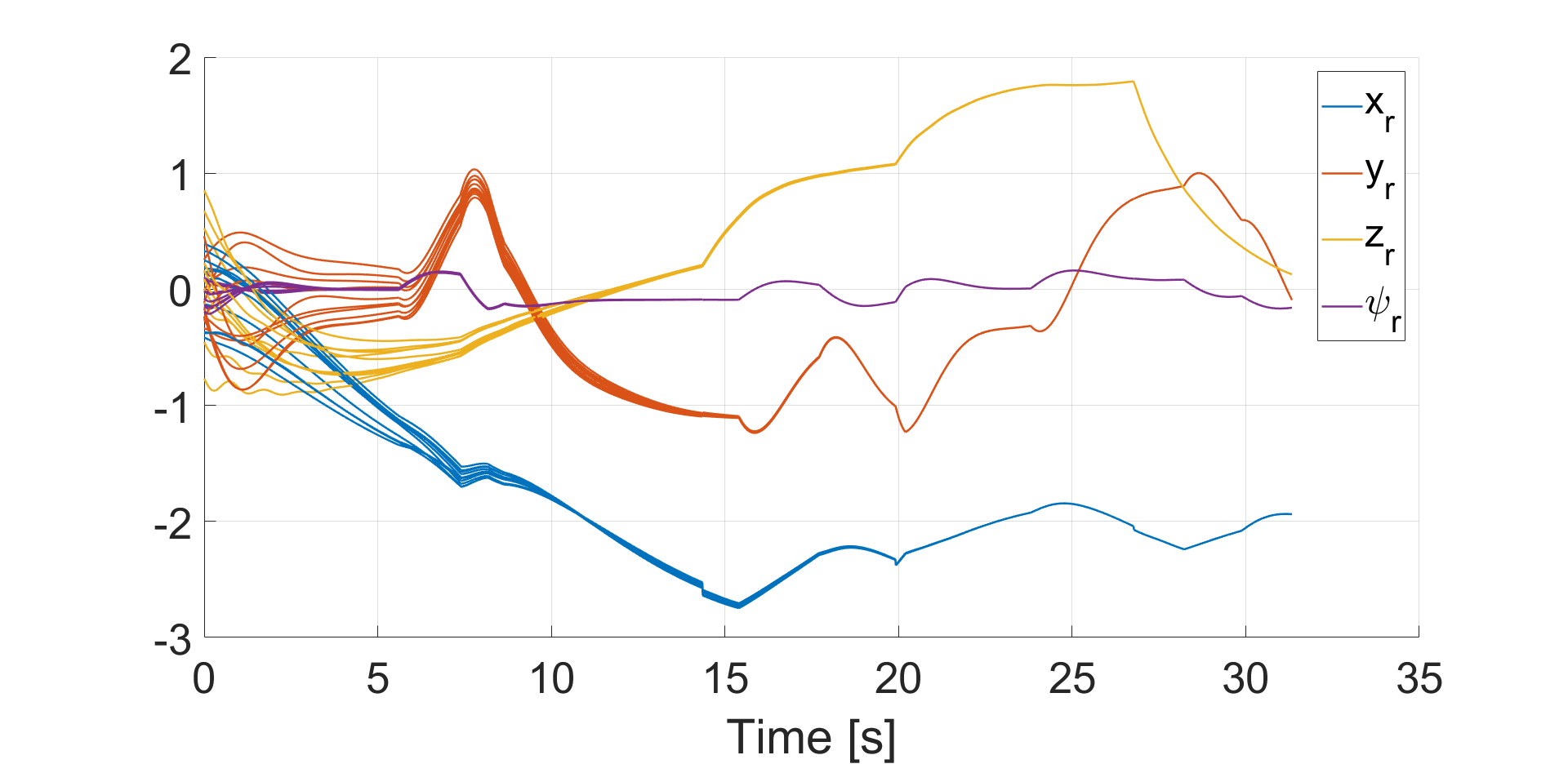}
		\caption{}
		\label{fig:plane_err_pos}
	\end{subfigure}\hspace{-0.5em}
	\begin{subfigure}[t]{0.48\textwidth}
		\centering
		\includegraphics[width=\textwidth,clip]{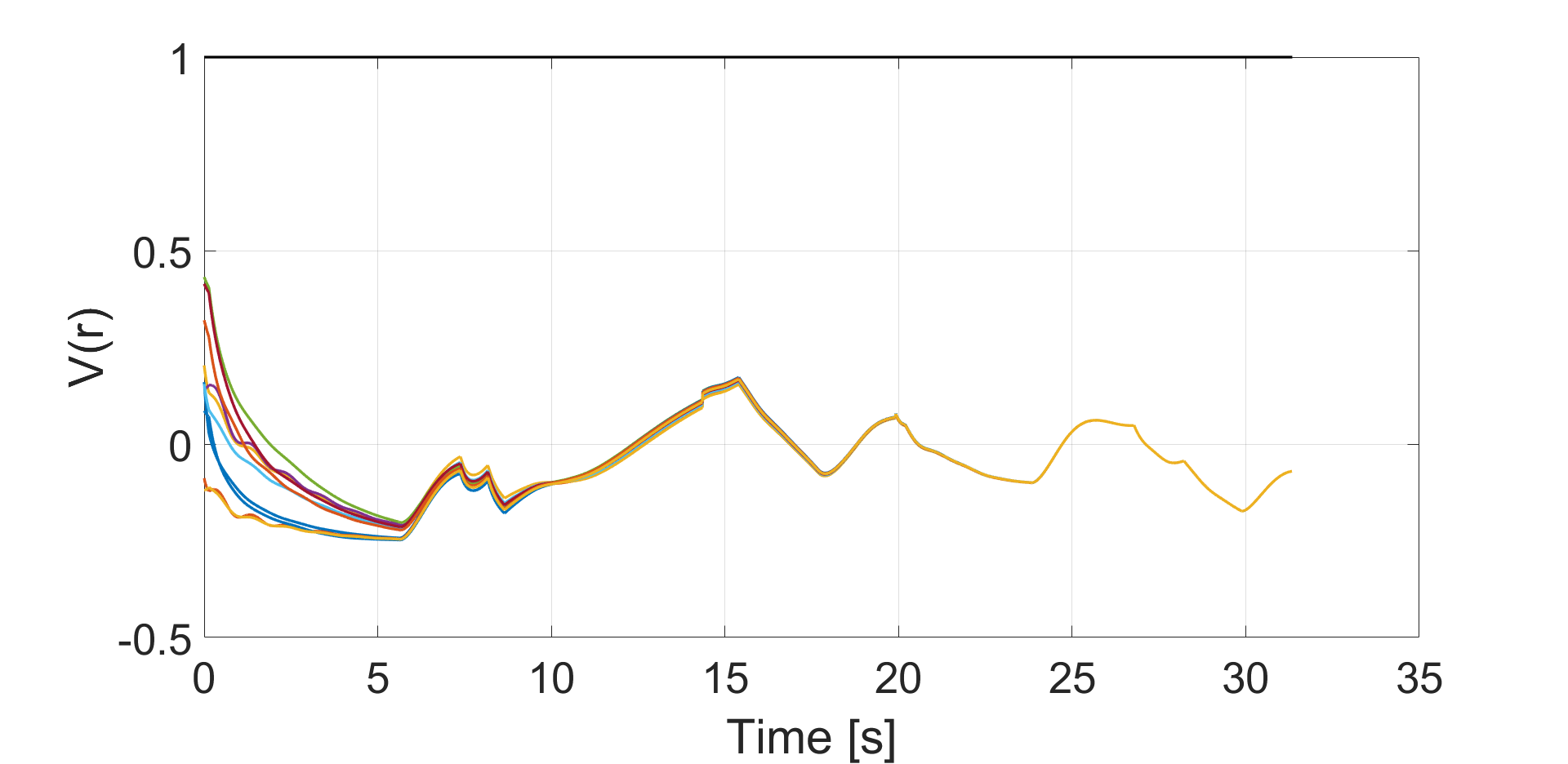}
		\caption{}
		\label{fig:plane_err_V}
	\end{subfigure}	
	\caption{(a) Error states $e = (x_r,y_r,z_r,\psi_r)$; (b) Verification of $\vf(\rstate(t)) \leq 1$ for all $t\geq 0$, for the nominal planned path in Figure~\ref{fig:plane_env} for a variety of initial conditions.}
	\label{fig:plane_err}
\end{figure}
}

\end{document}